\documentclass[a4paper,10pt]{article}
\usepackage{amssymb}
\usepackage{amsmath}
\usepackage{epsfig}
\usepackage{graphics}
\usepackage{subfig}
\usepackage{latexsym}
\usepackage{rotating}
\usepackage{titlesec}
\newcommand{\squeezeup}{\vspace{-9.5mm}}

\title{Higher order Hirota bilinear forms \thanks{A contribution to Metin G\"{u}rses' Festschrift (GURSES-FS-2025).}}

\author{Metin G\"{u}rses \thanks{gurses@fen.bilkent.edu.tr}\\
{\small Department of Mathematics, Faculty of Science}\\
{\small Bilkent University, 06800 Ankara - Turkey}\\
Asl{\i} Pekcan \thanks{Corresponding Author: aslipekcan@hacettepe.edu.tr} \\
{\small Department of Mathematics, Faculty of Science} \\
{\small Hacettepe University, 06800 Ankara - Turkey}
}

\date{\nonumber}
\begin{document}
\maketitle
\date{\nonumber}

\newtheorem{thm}{Theorem}[section]
\newtheorem{Le}{Lemma}[section]
\newtheorem{defi}{Definition}[section]
\newtheorem{ex}{Example}[section]
\newtheorem{pro}{Proposition}[section]
\baselineskip 17pt

\numberwithin{equation}{section}

\begin{abstract}
In this paper we study Hirota bilinear forms of the type $P(D) \{f\cdot f\}=0$.
We prove that for $P(D)=D_x^mD_y^rD_t^n$ the equations have three-soliton solutions if only if two of nonzero $m,n,p$ are odd and the other one even.
We explicitly derive the nonlinear partial differential equations corresponding to this form for $m+n+p=4$ and $m+n+p=6$.
We show that the equations for $P(D)=D_x(D_x^3+\alpha_1 D_t+\alpha_2 D_y)^{2k+1}$ possess three-soliton solutions for any constants $(\alpha_1,\alpha_2)\neq (0,0)$ and $k\in \mathbb{N}$. We conjecture that these equations have four-soliton solution only for $k=0$. Finally, we consider the equations for $P(D)=D_x^{m_1}D_y^{m_2}D_t^{m_3}D_z^{m_4}$. We prove that these equations have three-soliton solutions if only if one of $m_i=1$, and all the other $m_i$'s are odd for $i=1,2,3,4$. We observe that the monomials $D_x^mD_y^rD_t^n$ and $D_x^{m_1}D_y^{m_2}D_t^{m_3}D_z^{m_4}$ do not result genuine four-soliton solutions. In addition, we obtain three-soliton, lump, and hybrid solutions of these three type of equations for particular powers of the Hirota $D$-operators.

\end{abstract}


\tableofcontents

\section{Introduction}

Hirota bilinear method is a systematical way of finding soliton solutions of nonlinear integrable (even nonintegrable) partial
differential equations \cite{hir}. Soliton solutions of many well-known partial differential equations such as Korteweg-de Vries (KdV), modified KdV (MKdV), sine-Gordon equations had been derived by this method \cite{hir2}-\cite{hir4}. These type of solutions and their particular reductions like rogue, breather, lump wave solutions, play important
roles in real world problems; e.g. fiber optics \cite{fiber}, Bose-Einstein condensates \cite{bose}, Josephson junctions \cite{joseph}, plasma physics \cite{plasma} etc.

We have the following definition for the Hirota integrability \cite{hit1}.
\begin{defi} A set of equations written in the Hirota bilinear form is Hirota
integrable, if one can combine any number $N$ of one-soliton solutions into an
$N$-soliton solution, and the combination is a finite polynomial of exponential functions.
\end{defi}

One can construct one- and two-soliton solutions of all nonlinear partial differential equations including nonintegrable equations which can be written as
\begin{equation}
P(D_x,D_y,D_t,\ldots) \{f\cdot f\}=0.
\end{equation}
Here $P$ is an even polynomial of Hirota $D$-operator which is defined as
{\small
\begin{equation}
D_x^mD_y^rD_t^n\{f\cdot f\}=\Big(\frac{\partial}{\partial x}-\frac{\partial}{\partial x'}\Big)^m\Big(\frac{\partial}{\partial y}-\frac{\partial}{\partial y'}\Big)^r\Big(\frac{\partial}{\partial t}-\frac{\partial}{\partial t'}\Big)^n\,f(x,y,t)f(x',y',t')|_{x'=x,t'=t,y'=y},
\end{equation}}
where $m,r,n$ are positive integers and $x,y,t$ are independent variables.

The existence of three-soliton solution is very restrictive which is used as a sign of integrability. A nonlinear partial differential equation has three-soliton solution if it satisfies the three-soliton solution condition (3SC) given by
\begin{equation}
\sum_{\substack{1\le i<j\le 3 \\ \sigma_i = \pm 1}} P(\sigma_1 \vec{p}_1+\sigma_2\vec{p}_2+\sigma_3\vec{p}_3)P(\sigma_1\vec{p}_1-\sigma_2\vec{p}_2)P(\sigma_2\vec{p}_2-\sigma_3\vec{p}_3)P(\sigma_3\vec{p}_3-\sigma_1\vec{p}_1)=0,
\end{equation}
or explicitly
\begin{align}\label{3SC}
&P(\vec{p}_1-\vec{p}_2)P(\vec{p}_1-\vec{p}_3)P(\vec{p}_2-\vec{p}_3)P(\vec{p}_1+\vec{p}_2+\vec{p}_3)\nonumber\\
&+P(\vec{p}_2-\vec{p}_3)P(\vec{p}_1+\vec{p}_2)P(\vec{p}_1+\vec{p}_3)P(\vec{p}_1-\vec{p}_2-\vec{p}_3)\nonumber\\
&+P(\vec{p}_1-\vec{p}_3)P(\vec{p}_1+\vec{p}_2)P(\vec{p}_2+\vec{p}_3)P(\vec{p}_1-\vec{p}_2+\vec{p}_3)\nonumber\\
&+P(\vec{p}_1-\vec{p}_2)P(\vec{p}_1+\vec{p}_3)P(\vec{p}_2+\vec{p}_3)P(\vec{p}_1+\vec{p}_2-\vec{p}_3)=0,
\end{align}
in addition to the dispersion relation $P(\vec{p}_i)=0$, $i=1, 2, 3$. Here $\vec{p}_i$, $i=1, 2, 3$,  are the vectors of parameters. Hietarinta \cite{hit2}-\cite{hit6} used this condition as a tool to
search for integrable equations. Even a nonlinear partial differential equation possesses three-soliton solution, it is not obvious that it has $N$-soliton solution. For instance, when we check for four-soliton solution we again get a very restrictive condition, called the four-soliton solution condition (4SC)
\begin{align}\label{4SC}
&P(\vec{p}_1-\vec{p}_2)P(\vec{p}_1-\vec{p}_3)P(\vec{p}_1-\vec{p}_4)P(\vec{p}_2-\vec{p}_3)P(\vec{p}_2-\vec{p}_4)P(\vec{p}_3-\vec{p}_4)P(\vec{p}_1+\vec{p}_2+\vec{p}_3+\vec{p}_4)\nonumber\\
&-P(\vec{p}_2-\vec{p}_3)P(\vec{p}_2-\vec{p}_4)P(\vec{p}_3-\vec{p}_4)P(\vec{p}_1+\vec{p}_3)P(\vec{p}_1+\vec{p}_2)P(\vec{p}_1+\vec{p}_4)P(\vec{p}_1-\vec{p}_2-\vec{p}_3-\vec{p}_4)\nonumber\\
&-P(\vec{p}_1-\vec{p}_3)P(\vec{p}_1-\vec{p}_4)P(\vec{p}_3-\vec{p}_4)P(\vec{p}_1+\vec{p}_2)P(\vec{p}_2+\vec{p}_3)P(\vec{p}_2+\vec{p}_4)P(\vec{p}_1-\vec{p}_2+\vec{p}_3+\vec{p}_4)\nonumber\\
&-P(\vec{p}_1-\vec{p}_2)P(\vec{p}_1-\vec{p}_4)P(\vec{p}_2-\vec{p}_4)P(\vec{p}_1+\vec{p}_3)P(\vec{p}_2+\vec{p}_3)P(\vec{p}_3+\vec{p}_4)P(\vec{p}_1+\vec{p}_2-\vec{p}_3+\vec{p}_4)\nonumber\\
&-P(\vec{p}_1-\vec{p}_2)P(\vec{p}_1-\vec{p}_3)P(\vec{p}_2-\vec{p}_3)P(\vec{p}_1+\vec{p}_4)P(\vec{p}_2+\vec{p}_4)P(\vec{p}_3+\vec{p}_4)P(\vec{p}_1-\vec{p}_2+\vec{p}_3-\vec{p}_4)\nonumber\\
&+P(\vec{p}_1-\vec{p}_2)P(\vec{p}_3-\vec{p}_4)P(\vec{p}_1+\vec{p}_3)P(\vec{p}_1+\vec{p}_4)P(\vec{p}_2+\vec{p}_3)P(\vec{p}_2+\vec{p}_4)P(\vec{p}_1+\vec{p}_2-\vec{p}_3-\vec{p}_4)\nonumber\\
&+P(\vec{p}_1-\vec{p}_3)P(\vec{p}_2-\vec{p}_4)P(\vec{p}_1+\vec{p}_2)P(\vec{p}_1+\vec{p}_4)P(\vec{p}_2+\vec{p}_3)P(\vec{p}_3+\vec{p}_4)P(\vec{p}_1-\vec{p}_2+\vec{p}_3-\vec{p}_4)\nonumber\\
&+P(\vec{p}_1-\vec{p}_4)P(\vec{p}_2-\vec{p}_3)P(\vec{p}_1+\vec{p}_2)P(\vec{p}_1+\vec{p}_3)P(\vec{p}_2+\vec{p}_4)P(\vec{p}_3+\vec{p}_4)P(\vec{p}_1-\vec{p}_2-\vec{p}_3+\vec{p}_4)=0.
\end{align}

In this work we shall continue on the Hietarinta's work further by increasing the powers of the operators $D_{x}, D_{y}$, and $D_{t}$ in $(2+1)$- and $(3+1)$-dimensions.
Recently, we observe some interesting works on extending the works of Hietarinta. In particular Ma and his colleagues produced some examples of $(2+1)$-dimensional Hirota integrable partial differential equations \cite{Ma-0}-\cite{Lin}. The main idea in all these works is to obtain the soliton solutions of the associated nonlinear partial differential equations for a given Hirota bilinear form. Another important reason for studying Hirota bilinearization is to obtain not only soliton solutions by the Hirota method but also other kind of solutions such as the complexitons, positons, negatons, lump, and hybrid solutions of the nonlinear partial differential equations whose bilinearization is possible.

In this paper we study the Hirota bilinear form $D_x^mD_y^rD_t^n\{f\cdot f\}=0$ in Section 2. We obtain the conditions on the nonzero constants  $m,n,p$ for the existence of three-soliton solution. We present the nonlinear partial differential equations possessing three-soliton solutions corresponding to this form for $m+n+p=4$ and $m+n+p=6$ by the help of the equalities that we derive in Appendix. In Section 2, we also consider another $(2+1)$-dimensional Hirota bilinear form which is $D_x(D_x^3+\alpha_1 D_t+\alpha_2 D_y)^{2k+1}\{f\cdot f\}=0$. We prove that this new form satisfies the three-soliton solution condition for any constants $(\alpha_1,\alpha_2)\neq (0,0)$ and $k\in \mathbb{N}$. We then give a conjecture that these equations have four-soliton solution only for $k=0$. In Section 3, we analyze the $(3+1)$-dimensional Hirota bilinear form $D_x^{m_1}D_y^{m_2}D_t^{m_3}D_z^{m_4}\{f\cdot f\}=0$. We show that these equations possess three-soliton solutions if only if one of $m_i=1$, and all the other $m_i$'s are odd for $i=1,2,3,4$. In Section 4, we give particular examples of three-soliton solutions of these three Hirota bilinear equations mentioned in the previous sections. Among these equations, only equation $D_x(D_x^3+\alpha_1 D_t+\alpha_2 D_y)^{2k+1}\{f\cdot f\}=0$ for $k=0$ admits a four-soliton solution, therefore, we also include its four-soliton solution. In Section 5 and Section 6, we give lump and hybrid solutions of these equations for specific powers of the Hirota $D$-operators, respectively. At the end, we add an Appendix that contains fourth and sixth order bilinear forms.

\section{$(2+1)$-dimensional Hirota bilinear forms}

In this section we analyze two $(2+1)$-dimensional Hirota bilinear equations: $D_x^mD_y^nD_t^p\{f\cdot f\}=0$ and
$D_x(D_x^3+\alpha_1 D_t+\alpha_2 D_y)^{2k+1}\{f\cdot f\}=0$.

\subsection{Equations of the form $D_x^mD_y^nD_t^p\{f\cdot f\}=0$}

In \cite{hit2}, Hietarinta presented his work on searching Hirota bilinear forms in the form $P(D)\{f\cdot f\}=0$, which
have three-soliton solutions by the help of a symbolic computation language (REDUCE). In that work, in addition to the monomials in two and three variables, the polynomials $P(D)$ of degree up to eight had been considered. Hietarinta gave the conditions on the constants $m,n,p$ in $P(D)=D_x^mD_y^nD_t^p$ for the existence of three-soliton solution.
Here, we give a similar proof in detail.

\begin{Le}
The Hirota bilinear equation
\begin{equation}\label{Dx^mD_y^nD_t^p}
D_x^mD_y^nD_t^p\{f\cdot f\}=0,\, m+n+p=\mathrm{even},\, m, n, p \in \mathbb{N}
\end{equation}
 possesses three-soliton solutions if only if two of nonzero $m,n,p$ are odd and the other one even.
\end{Le}

\noindent \textbf{Proof.} Here we check the three-soliton solution condition (\ref{3SC}). To obtain three-soliton solution we let $f=1+\varepsilon f_1+\varepsilon^2 f_2+\varepsilon_3f_3$, $f_1=e^{\theta_1}+e^{\theta_2}+e^{\theta_3}$, $\theta_i=k_ix+\omega_it+l_iy+\delta_i$, $i=1,2,3$. Inserting this form into the Hirota bilinear form (\ref{Dx^mD_y^nD_t^p}) gives the dispersion relation
\begin{equation}
k_i^ml_i^n\omega_i^p=0,\quad i=1,2,3.
\end{equation}
To obtain nondegenerate phase shifts we take, for instance, $k_1=l_2=\omega_3=0$. In this case (\ref{3SC}) becomes
\begin{align}
(k_2k_3(k_2^2-k_3^2))^m(\omega_1\omega_2(\omega_1^2-\omega_2^2))^p(l_1l_3(l_1^2-l_3^2))^nA=0,
\end{align}
where
\begin{align}
A=&(-1)^{m+n+p}+(-1)^{p+n}+(-1)^{m+p}+(-1)^{n+m}\nonumber\\
=&1+(-1)^{-m}+(-1)^{-n}+(-1)^{-p}.
\end{align}
Obviously, $A=0$ that is the three-soliton solutions condition (\ref{3SC}) is satisfied if two of $m,n,p$ are odd and the other one even. Then we obtain the three-soliton solution of the equation (\ref{Dx^mD_y^nD_t^p}) as
\begin{align}\label{three-solDx^mD_y^nD_t^p}\displaystyle
u=&2[k_1e^{\theta_1}+k_2e^{\theta_2}+k_3e^{\theta_3}+A_{12}(k_1+k_2)e^{\theta_1+\theta_2}+A_{13}(k_1+k_3)e^{\theta_1+\theta_3}
+A_{23}(k_2+k_3)e^{\theta_2+\theta_3}\nonumber\\&+(k_1+k_2+k_3)A_{12}A_{13}A_{23}e^{\theta_1+\theta_2+\theta_3}]\Bigg / \nonumber\\
&\Big[1+e^{\theta_1}+e^{\theta_2}+e^{\theta_3}+A_{12}e^{\theta_1+\theta_2}+A_{13}e^{\theta_1+\theta_3}+A_{23}e^{\theta_2+\theta_3}+A_{12}A_{13}A_{23}e^{\theta_1+\theta_2+\theta_3}\Big],
\end{align}
where
\begin{equation}
A_{ij}=-\frac{P(\vec{p}_i-\vec{p}_j)}{P(\vec{p}_i+\vec{p}_j)}=-\frac{(k_i-k_j)^m(l_i-l_j)^n(\omega_i-\omega_j)^p}{k_i+k_j)^m(l_i+l_j)^n(\omega_i+\omega_j)^p}
\end{equation}
for $1\leq i<j\leq 3$, $\theta_1=\omega_1 t+l_1y+\delta_1$, $\theta_2=k_2x+\omega_2 t+\delta_2$, and $\theta_3=k_3x+l_3y+\delta_3$.

In the literature, the corresponding nonlinear partial differential equations passing three-soliton solution condition having Hirota
bilinear form (\ref{Dx^mD_y^nD_t^p}) were not given explicitly. Here we present these equations obtained by the bilinearizing transformation $u=2(\ln f)_x$ by utilizing the identities that we obtain in Appendix. Let us consider the cases when $m+n+p=4$ and $m+n+p=6$.  These equations are nonlocal, i.e., they are involving $D^{-1}$ terms. We can localize them by defining $u=w_x$, where $w$ is the new variable.

\noindent \textbf{Case 1. m+n+p=4}.

\noindent (i) $m=2, p=1, n=1$.
\begin{equation}
u_{xxyt}+u_xu_{yt}+2u_yu_{xt}+2u_tu_{xy}+u_{xx}(D^{-1}u_{yt})=0.
\end{equation}
\noindent Localization by $u=w_x$, then we have
\begin{equation}
w_{xxxyt}+w_{xx}\,w_{xyt}+2w_{xy}\,w_{xxt}+2\,w_{xt}\,w_{xxy}+w_{xxx}\,w_{yt}=\left(w_{xxyt}+w_{xx}\,w_{yt}+2 w_{xy}\, w_{xt}\right)_{x}=0.
\end{equation}

\noindent (ii) $m=1, p=2, n=1$.
\begin{equation}
u_{xytt}+u_yu_{tt}+2u_tu_{yt}+u_{xy}(D^{-1}u_{tt})+2u_{xt}(D^{-1}u_{yt})=0.
\end{equation}
\noindent Localization by $u=w_x$, then we have
\begin{equation}
w_{xxytt}+w_{xy}\,w_{xtt}+2\,w_{xt}\, w_{xyt}+w_{xxy}\,w_{tt}+2\,w_{xxt}\,w_{yt}=\left(w_{xytt}+w_{xy}\,w_{tt}+2 w_{ty}\, w_{xt}\right)_{x}=0
\end{equation}

\noindent (iii) $m=1, p=1, n=2$.
\begin{equation}
u_{xtyy}+u_tu_{yy}+2u_yu_{yt}+u_{xt}(D^{-1}u_{yy})+2u_{xy}(D^{-1}u_{yt})=0.
\end{equation}
\noindent Localization by $u=w_x$, then we have
\begin{equation}
w_{xxtyy}+w_{xt}\,w_{xyy}+2\,w_{xy}\,w_{xyt}+w_{xxt}\,w_{yy}+2\, w_{xxy}\,w_{yt}=\left(w_{xtyy}+w_{xt}\,w_{yy}+2 w_{xy}\, w_{yt}\right)_{x}=0.
\end{equation}

\noindent \textbf{Case 2. m+n+p=6}.

\noindent (i) $m=1, p=1, n=4$.
{\small\begin{align}
&u_{yyyyxt}+u_tu_{yyyy}+4u_{yt}u_{yyy}+6u_{yy}u_{tyy}+4u_yu_{tyyy}
+\Big(6u_tu_{yy}+12u_yu_{yt}+6u_{xtyy}
+3u_{xt}(D^{-1}u_{yy})\Big)(D^{-1}u_{yy})\nonumber\\&+4u_{xy}(D^{-1}u_{tyyy})
+u_{xt}(D^{-1}u_{yyyy})+\Big(4u_{xyyy}+12u_yu_{yy}+12u_{xy}(D^{-1}u_{yy})\Big)(D^{-1}u_{yt})=0.
\end{align}}
\noindent Localization by $u=w_x$, then we have
{\small \begin{align}
&w_{xxyyyyt}+w_{xt}w_{xyyyy}+4w_{xyt}w_{xyyy}+6w_{xyy}w_{xtyy}+4w_{xy}w_{xtyyy}+\Big(6w_{xt}w_{xyy}+12w_{xy}w_{xyt}+6w_{xxtyy}\nonumber\\
&+3w_{xxtw_{yy}}\Big)w_{yy}
+4w_{xxy}w_{tyyy}
+w_{xxt}w_{yyyy}
+\Big(4w_{xxyyy}+12w_{xy}w_{xyy}+12w_{xxy}w_{yy}\Big)w_{yt}\nonumber\\
&=(w_{xyyyyt}+4w_{xy}w_{tyyy}+6w_{xtyy}w_{yy}+12w_{xy}w_{yy}w_{yt}+w_{xt}w_{yyyy}+4w_{yt}w_{xyyy}+3w_{xt}w_{yy}^2)_x=0.
\end{align}}

\noindent (ii) $m=2, p=1, n=3$.
{\small \begin{align}
&u_{yyyxxt}+u_xu_{yyyt}+6u_yu_{xyyt}+6u_y^2u_{yt}+2u_tu_{xyyy}+6u_tu_yu_{yy}+6u_{xy}u_{yyt}+3u_{yt}u_{xyy}+2u_{yyy}u_{xt}+3u_{yy}u_{xyt}
\nonumber\\
&+\Big(3u_xu_{yt}+6u_yu_{xt}+6u_tu_{xy}+3u_{xxyt}\Big)(D^{-1}u_{yy})+\Big(3u_{xxyy}+12u_yu_{xy}+3u_xu_{yy}+3u_{xx}(D^{-1}u_{yy})\Big)(D^{-1}u_{yt})
\nonumber\\
&+u_{xx}(D^{-1}u_{yyyt})=0.
\end{align}}
\noindent Localization by $u=w_x$, then we have
{\small \begin{align}
&w_{yyyxxxt}+w_{xx}w_{xyyyt}+6w_{xy}w_{xxyyt}+6w_{xy}^2w_{xyt}+2w_{xt}w_{xxyyy}+6w_{xt}w_{xy}w_{xyy}+6w_{xxy}w_{xyyt}+3w_{xyt}w_{xxyy}
\nonumber\\
&+2w_{xyyy}w_{xxt}+3w_{xyy}w_{xxyt}+\Big(3w_{xx}w_{xyt}+6w_{xy}w_{xxt}+6w_{xt}w_{xxy}+3w_{xxxyt}\Big)w_{yy}+w_{xxx}w_{yyyt}\nonumber\\
&+\Big(3w_{xxxyy}+12w_{xy}w_{xxy}+3w_{xx}w_{xyy}+3w_{xxx}w_{yy}\Big)w_{yt}\nonumber\\
&=(w_{yyyxxt}+w_{xx}w_{yyyt}+6w_{xy}^2w_{yt}+2w_{xt}w_{xyyy}+6w_{xt}w_{xy}w_{yy}+6w_{xy}w_{xyyt}+3w_{xxyt}w_{yy}\nonumber\\
&+3w_{xx}w_{yt}w_{yy}+3w_{yt}w_{xxyy})_x=0.
\end{align}}

\noindent (iii) $m=3, p=1, n=2$.
{\small\begin{align}\label{m=3p=1n=2}
&u_{xxxyyt}+6u_{xy}u_{xyt}+3u_{xx}u_{yyt}+2u_{yt}u_{xxy}+3u_{xt}u_{xyy}+6u_yu_{xxyt}+3u_xu_{xyyt}+6u_{xt}u_y^2+3u_tu_{xxyy}
+3u_xu_tu_{yy}\nonumber\\
&+u_{yy}u_{xxt}+12u_tu_yu_{xy}+6u_xu_yu_{yt}+\Big(2u_{xxxy}+6u_xu_{xy}+6u_yu_{xx}\Big)(D^{-1}u_{yt})\nonumber\\
&+\Big(u_{xxxt}+3u_xu_{xt}
+3u_tu_{xx}\Big)(D^{-1}u_{yy})=0.
\end{align}}
\noindent  Localization by $u=w_x$, then we have
{\small\begin{align}
&w_{xxxxyyt}+6w_{xxy}w_{xxyt}+3w_{xxx}w_{xyyt}+2w_{xyt}w_{xxxy}+3w_{xxt}w_{xxyy}+6w_{xy}w_{xxxyt}+3w_{xx}w_{xxyyt}+6w_{xxt}w_{xy}^2
\nonumber\\
&+3w_{xt}w_{xxxyy}
+3w_{xx}w_{xt}w_{xyy}+w_{xyy}w_{xxxt}+12w_{xt}w_{xy}w_{xxy}+6w_{xx}w_{xy}w_{xyt}
\nonumber\\
&+\Big(2w_{xxxxy}+6w_{xx}w_{xxy}+6w_{xy}w_{xxx}\Big)w_{yt}+\Big(w_{xxxxt}+3w_{xx}w_{xxt}
+3w_{xt}w_{xxx}\Big)w_{yy}\nonumber\\
&=(w_{xxxyyt}+6w_{xy}w_{xxyt}+6w_{xt}w_{xy}^2+w_{yy}w_{xxxt}+2w_{yt}w_{xxxy}+3w_{xx}w_{xyyt}+6w_{xx}w_{xy}w_{yt}\nonumber\\
&+3w_{xx}w_{xt}w_{yy}+3w_{xt}w_{xxyy})_x=0.
\end{align}}

\noindent (iv) $m=4, p=1, n=1$.
{\small \begin{align}
&u_{xxxxyt}+12u_tu_yu_{xx}+u_{xxx}u_{yt}+12u_xu_yu_{xt}+12u_xu_tu_{xy}+4u_{xy}u_{xxt}+6u_{xx}u_{xyt}
+4u_{xt}u_{xxy}+4u_yu_{xxxt}\nonumber\\
&+6u_xu_{xxyt}+3u_x^2u_{yt}+4u_tu_{xxxy}+\Big(u_{xxxx}+6u_xu_{xx}\Big)(D^{-1}u_{yt})=0.
\end{align}}
\noindent Localization by $u=w_x$, then we have
{\small \begin{align}
&w_{xxxxxyt}+12w_{xt}w_{xy}w_{xxx}+w_{xxxx}w_{xyt}+12w_{xx}w_{xy}w_{xxt}+12w_{xx}w_{xt}w_{xxy}+4w_{xxy}w_{xxxt}+6w_{xxx}w_{xxyt}
\nonumber\\
&+4w_{xxt}w_{xxxy}+4w_{xy}w_{xxxxt}+6w_{xx}w_{xxxyt}+3w_{xx}^2w_{xyt}+4w_{xt}w_{xxxxy}+\Big(w_{xxxxx}+6w_{xx}w_{xxx}\Big)w_{yt}\nonumber\\
&=(w_{xxxxyt}+3w_{xx}^2w_{yt}+w_{xxxx}w_{yt}+4w_{xy}w_{xxxt}+4w_{xt}w_{xxxy}+12w_{xt}w_{xy}w_{xx}+6w_{xx}w_{xxyt})_x=0.
\end{align}}

\noindent (v) $m=1, p=2, n=3$.
{\small \begin{align}
&u_{yyyxtt}+3u_{ytt}u_{yy}+6u_{yt}u_{yyt}+3u_yu_{yytt}+2u_tu_{yyyt}+u_{tt}u_{yyy}+\Big(u_{xyyy}+3u_yu_{yy}+3u_{xy}(D^{-1}u_{yy})\Big)(D^{-1}u_{tt})
\nonumber\\
&+\Big(3u_{xytt}+3u_{y}u_{tt}+6u_tu_{yt}+6u_{xt}(D^{-1}u_{yt})\Big)(D^{-1}u_{yy})+3u_{xy}(D^{-1}u_{yytt})+2u_{xt}(D^{-1}u_{yyyt})\nonumber\\
&+\Big(6u_{xyyt}+6u_tu_{yy}+12u_yu_{yt}+6u_{xy}(D^{-1}u_{yt})\Big)(D^{-1}u_{yt})=0.
\end{align}}
\noindent  Localization by $u=w_x$, then we have
{\small \begin{align}
&w_{yyyxxtt}+3w_{xytt}w_{xyy}+6w_{xyt}w_{xyyt}+3w_{xy}w_{xyytt}+2w_{xt}w_{xyyyt}+w_{xtt}w_{xyyy}+3w_{xxy}w_{yytt}+2w_{xxt}w_{yyyt}\nonumber\\
&+\Big(w_{xxyyy}+3w_{xy}w_{xyy}+3w_{xxy}w_{yy}\Big)w_{tt}
+\Big(3w_{xxytt}+3w_{xy}w_{xtt}+6w_{xt}w_{xyt}+6w_{xxt}w_{yt}\Big)w_{yy}\nonumber\\
&+\Big(6w_{xxyyt}+6w_{xt}w_{xyy}+12w_{xy}w_{xyt}+6w_{xxy}w_{yt}\Big)w_{yt}\nonumber\\
&=(w_{yyyxtt}+6w_{xy}w_{yt}^2+w_{tt}w_{xyyy}+2w_{xt}w_{yyyt}+3w_{xy}w_{tt}w_{yy}+3w_{xytt}w_{yy}+3w_{xy}w_{yytt} \nonumber\\
&+6w_{xt}w_{yt}w_{yy}+6w_{xyyt}w_{yt})_x=0.
\end{align}}

\noindent (vi) $m=1, p=4, n=1$.
{\small \begin{align}
&u_{ttttxy}+u_yu_{tttt}+4u_{yt}u_{ttt}+6u_{tt}u_{ytt}+4u_tu_{yttt}+u_{xy}(D^{-1}u_{tttt})+\Big(4u_{xttt}+12u_tu_{tt}+12u_{xt}(D^{-1}u_{tt})\Big)(D^{-1}u_{yt})\nonumber\\
&+\Big(6u_yu_{tt}+12u_tu_{yt}+6u_{xytt}+3u_{xy}(D^{-1}u_{tt})\Big)(D^{-1}u_{tt})+4u_{xt}(D^{-1}u_{yttt})=0.
\end{align}}
\noindent  Localization by $u=w_x$, then we have
{\small \begin{align}
&w_{ttttxxy}+w_{xy}w_{xtttt}+4w_{xyt}w_{xttt}+6w_{xtt}w_{xytt}+4w_{xt}w_{xyttt}+w_{xxy}w_{tttt}+4w_{xxt}w_{yttt}\nonumber\\
&+\Big(6w_{xy}w_{xtt}+12w_{xt}w_{xyt}+6w_{xxytt}+3w_{xxy}w_{tt}\Big)w_{tt}+\Big(4w_{xxttt}+12w_{xt}w_{xtt}+12w_{xxt}w_{tt}\Big)w_{yt}\nonumber\\
&=(w_{ttttxy}+w_{xy}w_{tttt}+4w_{xt}w_{yttt}+4w_{yt}w_{xttt}+3w_{xy}w_{tt}^2+12w_{xt}w_{tt}w_{yt}+6w_{tt}w_{xytt})_x=0.
\end{align}}

\noindent (vii) $m=1, p=3, n=2$.
{\small \begin{align}
&u_{tttxyy}+3u_{tyy}u_{tt}+6u_{yt}u_{tty}+3u_tu_{yytt}+2u_yu_{ttty}+u_{yy}u_{ttt}+\Big(u_{xttt}+3u_tu_{tt}+3u_{xt}(D^{-1}u_{tt})\Big)(D^{-1}u_{yy})\nonumber\\
&+\Big(3u_{xtyy}+3u_{t}u_{yy}+6u_yu_{yt}+6u_{xy}(D^{-1}u_{yt})\Big)(D^{-1}u_{tt})+3u_{xt}(D^{-1}u_{yytt})+2u_{xy}(D^{-1}u_{ttty})\nonumber\\
&+\Big(6u_{xtty}+6u_yu_{tt}+12u_tu_{yt}+6u_{xt}(D^{-1}u_{yt})\Big)(D^{-1}u_{yt})=0.
\end{align}}
\noindent  Localization by $u=w_x$, then we have
{\small \begin{align}
&w_{tttxxyy}+3w_{xtyy}w_{xtt}+6w_{xyt}w_{xtty}+3w_{xt}w_{xyytt}+2w_{xy}w_{xttty}+w_{xyy}w_{xttt}+\Big(w_{xxttt}+3w_{xt}w_{xtt}+3w_{xxt}w_{tt}\Big)w_{yy}\nonumber\\
&+\Big(3w_{xxtyy}+3w_{xt}w_{xyy}+6w_{xy}w_{xyt}+6w_{xxy}w_{yt}\Big)w_{tt}+3w_{xxt}w_{yytt}
+2w_{xxy}w_{ttty}\nonumber\\
&+\Big(6w_{xxtty}+6w_{xy}w_{xtt}+12w_{xt}w_{xyt}+6w_{xxt}w_{yt}\Big)w_{yt}\nonumber\\
&=(w_{tttxyy}+w_{yy}w_{xttt}+6w_{xt}w_{yt}^2+2w_{xy}w_{ttty}+3w_{xt}w_{yytt}+3w_{xt}w_{tt}w_{yy}+6w_{xy}w_{yt}w_{xtty}\nonumber\\
&+6w_{yt}w_{xtty}+3w_{xtyy}w_{tt})_x=0.
\end{align}}

\noindent (viii) $m=2, p=3, n=1$.
{\small \begin{align}
&u_{tttxxy}+u_xu_{ttty}+6u_tu_{xtty}+6u_t^2u_{yt}+2u_yu_{xttt}+6u_tu_yu_{tt}+3u_{tt}u_{xyt}+6u_{xt}u_{tty}+3u_{yt}u_{xtt}+2u_{ttt}u_{xy}
\nonumber\\
&+\Big(3u_xu_{yt}+6u_tu_{xy}+3u_{xxyt}+6u_yu_{xt}\Big)(D^{-1}u_{tt})+u_{xx}(D^{-1}u_{ttty})\nonumber\\
&+\Big(3u_{xxtt}+3u_xu_{tt}+12u_tu_{xt}+3u_{xx}(D^{-1}u_{tt})\Big)(D^{-1}u_{yt})=0.
\end{align}}
\noindent Localization by $u=w_x$, then we have
{\small \begin{align}
&w_{tttxxxy}+w_{xx}w_{xttty}+6w_{xt}w_{xxtty}+6w_{xt}^2w_{xyt}+2w_{xy}w_{xxttt}+6w_{xt}w_{xy}w_{xtt}+3w_{xtt}w_{xxyt}+w_{xxx}w_{ttty}\nonumber\\
&+6w_{xxt}w_{xtty}+3w_{xyt}w_{xxtt}+2w_{xttt}w_{xxy}+\Big(3w_{xx}w_{xyt}+6w_{xt}w_{xxy}+3w_{xxxyt}+6w_{xy}w_{xxt}\Big)w_{tt}\nonumber\\
&+\Big(3w_{xxxtt}+3w_{xx}w_{xtt}+12w_{xt}w_{xxt}+3w_{xxx}w_{tt}\Big)w_{yt}\nonumber\\
&=(w_{tttxxy}+w_{xx}w_{ttty}+6w_{xt}^2w_{yt}+2w_{xy}w_{xttt}+3w_{xx}w_{yt}w_{tt}+6w_{xt}w_{xy}w_{tt}+6w_{xt}w_{xtty}\nonumber\\
&+3w_{tt}w_{xxyt}+3w_{yt}w_{xxtt})_x=0.
\end{align}}

\noindent (ix) $m=3, p=2, n=1$.
{\small \begin{align}
&u_{xxxtty}+6u_{xt}u_{xyt}+3u_{xx}u_{tty}+2u_{yt}u_{xxt}+3u_{xy}u_{xtt}+6u_tu_{xxyt}+3u_xu_{xtty}+6u_{xy}u_t^2+3u_yu_{xxtt}
\nonumber\\
&+3u_xu_yu_{tt}+u_{tt}u_{xxy}+12u_tu_yu_{xt}+6u_xu_tu_{yt}+\Big(2u_{xxxt}+6u_xu_{xt}+6u_tu_{xx}\Big)(D^{-1}u_{yt})
\nonumber\\
&+\Big(u_{xxxy}+3u_xu_{xy}+3u_yu_{xx}\Big)(D^{-1}u_{tt})=0.
\end{align}}
\noindent  Localization by $u=w_x$, then we have
{\small \begin{align}
&w_{xxxxtty}+6w_{xxt}w_{xxyt}+3w_{xxx}w_{xtty}+2w_{xyt}w_{xxxt}+3w_{xxy}w_{xxtt}+6w_{xt}w_{xxxyt}+3w_{xx}w_{xxtty}+6w_{xxy}w_{xt}^2
\nonumber\\
&+3w_{xy}w_{xxxtt}+3w_{xx}w_{xy}w_{xtt}+w_{xtt}w_{xxxy}+12w_{xt}w_{xy}w_{xxt}+6w_{xx}w_{xt}w_{xyt}\nonumber\\
&+\Big(2w_{xxxxt}+6w_{xx}w_{xxt}+6w_{xt}w_{xxx}\Big)w_{yt}+\Big(w_{xxxxy}+3w_{xx}w_{xxy}+3w_{xy}w_{xxx}\Big)w_{tt}\nonumber\\
&=(w_{xxxtty}+6w_{xy}w_{xt}^2+w_{tt}w_{xxxy}+3w_{xx}w_{xy}w_{tt}+6w_{xt}w_{xx}w_{yt}+2w_{yt}w_{xxxt}+6w_{xt}w_{xxyt}\nonumber\\
&+3w_{xx}w_{xtty}+3w_{xy}w_{xxtt})_x=0.
\end{align}}

\noindent \textbf{Remark 1.} The Hirota bilinear equation (\ref{Dx^mD_y^nD_t^p}) does not possess a genuine four-soliton solution. The dispersion relation
forces at least one of the components $(k_i, l_i, \omega_i)$ to vanish for each soliton. For any four-soliton configuration at least two solitons
necessarily lie on the same coordinate plane, which gives $P(\vec{p}_i-\vec{p}_j)=P(\vec{p}_i+\vec{p}_j)=0$ for some $(i,j)$. Hence the corresponding
phase shift $A_{ij}=-\frac{P(\vec{p}_i-\vec{p}_j)}{P(\vec{p}_i-\vec{p}_j)}$ becomes indeterminate and only degenerate solution can be obtained.

\subsection{Equations of the form $D_x(D_x^3+\alpha_1 D_t+\alpha_2 D_y)^{2k+1}\{f\cdot f\}=0$}

In this part, we consider the equations in the form $D_x(D_x^3+\alpha_1 D_t+\alpha_2 D_y)^{2k+1}\{f\cdot f\}=0$ for $k\in \mathbb{N}$ and
$\alpha_1, \alpha_2$ constants. We have the following lemma.

\begin{Le}
The Hirota bilinear equation
\begin{equation}\label{bilinearProb2}
D_x(D_x^3+\alpha_1 D_t+\alpha_2 D_y)^{2k+1}\{f\cdot f\}=0,\quad k=0,1,2,\ldots
\end{equation}
possesses three-soliton solutions for any constants $(\alpha_1,\alpha_2)\neq (0,0)$ and $k\in \mathbb{N}$.
\end{Le}

\noindent \textbf{Proof.} Consider the three-soliton solution condition (\ref{3SC}) where $P(D)=D_x(D_x^3+\alpha_1 D_t+\alpha_2 D_y)^{2k+1}$, $k=0,1,2,\ldots\,$  for any constants $(\alpha_1,\alpha_2)\neq (0,0)$. To derive three-soliton solution we take $f=1+\varepsilon f_1+\varepsilon^2 f_2+\varepsilon_3f_3$, $f_1=e^{\theta_1}+e^{\theta_2}+e^{\theta_3}$, $\theta_i=k_ix+\omega_it+l_iy+\delta_i$, $i=1,2,3$. Inserting this form into the Hirota bilinear equation (\ref{bilinearProb2}) we obtain the dispersion relation
\begin{equation}\label{dispPRO2}\displaystyle
\omega_i=-\frac{k_i^3+\alpha_2 l_i}{\alpha_1},\quad i=1,2,3,\quad \alpha_1\neq 0.
\end{equation}
We observe that
\begin{equation}
P(\sigma_i\vec{p}_i-\sigma_j\vec{p}_j)=-3^{2k+1}\sigma_i\sigma_j(k_ik_j)^{2k+1}(k_i-\sigma_i\sigma_jk_j)^{2k+2},
\end{equation}
and
{\small \begin{equation}
P(\sigma_1\vec{p}_1+\sigma_2\vec{p}_2+\sigma_3\vec{p}_3)=3^{2k+1}(\sigma_1k_1+\sigma_2k_2)^{2k+1}(\sigma_1k_1+\sigma_3k_3)^{2k+1}(\sigma_2k_2+\sigma_3k_3)^{2k+1}
(\sigma_1k_1+\sigma_2k_2+\sigma_3k_3),
\end{equation}}
for $\sigma_i=\pm 1$, $i=1, 2, 3$. Then under the above findings, the three-soliton solution condition (\ref{3SC}) becomes
\begin{equation}
-3^{4(2k+1)}(k_1k_2k_3)^{2(2k+1)}[(k_1^2-k_2^2)(k_1^2-k_3^2)(k_2^2-k_3^2)]^{(2k+1)} \Psi=0,
\end{equation}
where
\begin{align}
\Psi=&(k_1+k_2+k_3)(k_1-k_2)(k_2-k_3)(k_3-k_1)-(k_1+k_2)(k_1+k_3)(k_2-k_3)(-k_1+k_2+k_3)\nonumber\\
&-(k_1+k_2)(k_2+k_3)(k_3-k_1)(k_1-k_2+k_3)-(k_1+k_3)(k_2+k_3)(k_1-k_2)(k_1+k_2-k_3).
\end{align}
It is straightforward to show that $\Psi=0$ identically for any $k_1, k_2, k_3$. This means that 3SC is satisfied for any constants $(\alpha_1,\alpha_2)\neq (0,0)$. Three-soliton solution of the
equation (\ref{bilinearProb2}) for any $k\in \mathbb{N}$ is given by (\ref{three-solDx^mD_y^nD_t^p}) where
\begin{equation}\label{Prob2A_ij}
A_{ij}=\Big(\frac{k_i-k_j}{k_i+k_j}\Big)^{2k+2},\quad 1\leq i,j \leq 3,\quad k\in \mathbb{N},
\end{equation}
and $\theta_i=k_ix+\omega_i t+l_iy+\delta_i$ with $\omega_i=-\frac{k_i^3+\alpha_2 l_i}{\alpha_1}$ for $i=1, 2, 3$.

\noindent \textbf{Remark 2.} We note that $D_x(D_x^3+\alpha_1 D_t+\alpha_2 D_y+\alpha_3 D_z)^{2k+1}\{f\cdot f\}=0$, $k=0,1,2,\ldots\,$ also has
three-soliton solution for any $k$. In this case, by using the change of variables $X=x, T=t, Y=\alpha_2y+\alpha_3z$, $Z=-\alpha_3y+\alpha_2z$, the $(3+1)$-dimensional
form reduces to $(2+1)$-dimensional equation $D_X(D_X^3+\alpha_1D_T+D_Y)\{g\cdot g\}$, $f(x,y,z,t)=g(X,Y,T)$. It can be further transformed into $(1+1)$-dimensional equation. Indeed, more general equation $(D_x+\alpha_4D_y+a_5D_z)(D_x^3+\alpha_1 D_t+\alpha_2 D_y+\alpha_3 D_z)^{2k+1}\{f\cdot f\}=0$, $k=0,1,2,\ldots\,$ possess three-soliton solution for any $k$. This equation can also be reduced to $(1+1)$-dimensional equation $D_{\xi}(D_{\xi}^3+a_1D_t+(a_2a_4+a_3a_5)D_{\xi})\{g\cdot g\}=0$ by $\xi=x+a_4 y+a_5 z$. Here $f(x,y,z,t)=g(\xi,t)$.

In Section 4.2 we shall consider three-soliton solutions for the cases $k=0$ and $k=1$ of (\ref{bilinearProb2}). For $k=0$ we have
\begin{equation}
D_x(D_x^3+\alpha_1 D_t+\alpha_2 D_y)\{f\cdot f\}=0
\end{equation}
corresponding to
\begin{equation} \label{exk=0}
u_{xxxx}+6u_xu_{xx}+\alpha_1u_{xt}+\alpha_2u_{xy}=(u_{xxx}+3u_x^2+\alpha_1u_t+\alpha_2u_y)_x=0.
\end{equation}
\noindent For $k=1$ we have
\begin{equation}
D_x(D_x^3+\alpha_1 D_t+\alpha_2 D_y)^3\{f\cdot f\}=0.
\end{equation}
Since the equation corresponding to the above Hirota bilinear form is lengthy, we assume that $\alpha_2=0$, $f=f(x,t)$, and get the following $(1+1)$-dimensional
nonlinear partial differential equation for the function $u(x,t)$ as
\begin{align}\label{exk=1}
&u_{10x}+45u_xu_{8x}+45u_{xx}u_{7x}+210u_{xxx}u_{6x}+210u_{4x}u_{5x}+1575u_{xx}u_{xxx}^2+1260u_xu_{xx}u_{5x}\nonumber\\
&+3150u_xu_{xxx}u_{4x}+630u_x^2u_{6x}+9450u_x^2u_{xx}u_{xxx}+3150u_x^3u_{4x}+4725u_x^4u_{xx}\nonumber\\
&+3\alpha_1\Big(u_{7x,t}+7u_tu_{6x}+21u_xu_{5x,t}+7u_{xt}u_{5x}+21u_{xx}u_{4x,t}+35u_{xxt}u_{4x}+35u_{xxxt}u_{xxx}\nonumber\\
&+210u_xu_{xx}u_{xxt}+105u_xu_{xt}u_{xxx}+105u_x^2u_{xxxt}+105u_tu_{xx}u_{xxx}+105u_xu_tu_{4x}+105u_x^3u_{xt}+315u_tu_x^2u_{xx}\Big)\nonumber\\
&+3\alpha_1^2\Big(u_{xxxxtt}+6u_xu_{xxtt}+6u_{xx}u_{xtt}+8u_{xt}u_{xxt}+8u_{t}u_{xxxt}+u_{xxx}u_{tt}+3u_x^2u_{tt}+24u_xu_tu_{xt}
\nonumber\\
&+12u_{xx}u_t^2+(6u_xu_{xx}+u_{xxxx})(D^{-1}u_{tt})\Big)+\alpha_1^3\Big(u_{tttx}+3u_tu_{tt}+3u_{xt} (D^{-1} u_{tt})\Big)
=0.
\end{align}

\noindent \textbf{Remark 3.} We observe that the equation (\ref{bilinearProb2}) has four-soliton solution for $k=0$ which is given by $u=2(\ln(f))_x$ where
\begin{align}\label{Prob24SS}\displaystyle
f=1+\sum_{i=1}^4 e^{\theta_i}+\sum_{1\leq i< j\leq 4}A_{ij}e^{\theta_i+\theta_j}+\sum_{1\leq i< j<m\leq 4} A_{ijm}e^{\theta_i+\theta_j+\theta_m}+A_{12}A_{13}A_{14}A_{23}A_{24}A_{34}e^{\theta_1+\theta_2+\theta_3+\theta_4},
\end{align}
for $A_{ijm}=A_{ij}A_{jm}A_{im}$, $A_{ij}=\frac{(k_i-k_j)^2}{(k_i+k_j)^2}$, and $\omega_i=-\frac{k_i^3+\alpha_2 l_i}{\alpha_1}$, $i=1, 2, 3, 4$. Indeed, in \cite{ma2} and \cite{ma3}, it was shown that the equation (\ref{exk=0}) has $N$-soliton solution. For $k=1$, four-soliton solution condition (\ref{4SC}) is not satisfied directly. The condition becomes
\begin{align}
&-502096953744k_1^{11}k_2^{11}k_3^{11}k_4^{11}(k_3-k_4)^2(k_3+k_4)^2(k_2-k_4)^2(k_2+k_4)^2(k_2-k_3)^2
(k_2+k_3)^2(k_1-k_4)^2\nonumber\\
&\times(k_1+k_4)^2(k_1-k_3)^2(k_1+k_3)^2(k_1-k_2)^2(k_1+k_2)^2(k_1^2+k_2^2+k_3^2+k_4^2)=0.
\end{align}
To satisfy the above equality we need to put an additional condition on the solution parameters besides the dispersion relation which does not give genuine four-soliton solution.
We have the similar results for $k=2$ and $k=3$ hence we conjecture that the following lemma is valid for all $k$.
\begin{Le}
The equation (\ref{bilinearProb2}) has four-soliton solution only for $k=0$. For $k\geq 1$, it does not satisfy the four-soliton solution condition (\ref{4SC}) directly.
\end{Le}

\section{$(3+1)$-dimensional Hirota bilinear forms}

We shall now extend our analysis to  $(3+1)$-dimensional Hirota bilinear forms.

\subsection{Equations of the form $D_x^{m_1}D_y^{m_2}D_t^{m_3}D_z^{m_4}\{f\cdot f\}=0$}

In \cite{hir}, Hirota claims that by private communication with J. Hietarinta except the case
\begin{equation}\label{bilinearDxDyDtDz}
P(D)\{f\cdot f\}=D_xD_yD_tD_z\{f\cdot f\}=0,
\end{equation}
there appear to be no Hirota
bilinear equation having three-soliton solution but no four-soliton solution. As far as we know the corresponding nonlinear partial differential equation has been never given in the literature. Letting $u(x,y,t,z)=2(\ln f(x,y,t,z))_x$ we obtain the following nonlocal nonlinear partial differential equation
\begin{equation}\label{eqnDxDyDtDz}
u_{xytz}+u_yu_{zt}+u_zu_{yt}+u_tu_{yz}+u_{xt}D^{-1} (u_{zy})+u_{xz}D^{-1} (u_{yt})+u_{xy}D^{-1} (u_{zt})=0,\quad \Big(  D^{-1}=\int^x \Big).
\end{equation}

Let us study the generalized form of the equation (\ref{bilinearDxDyDtDz}).

\begin{Le}
The Hirota bilinear equation
\begin{equation}\label{Dx^m1D_y^m2D_t^m3D_z^m4}
D_x^{m_1}D_y^{m_2}D_t^{m_3}D_z^{m_4}\{f\cdot f\}=0,\, m_1+m_2+m_3+m_4=\mathrm{even},\, m_i\in \mathbb{N}
\end{equation}
 possesses three-soliton solutions if only if one of $m_i=1$, and all the other $m_i$'s are odd for $i=1,2,3,4$.
\end{Le}

\noindent \textbf{Proof.} To obtain three-soliton solution we insert $f=1+\varepsilon f_1+\varepsilon^2 f_2+\varepsilon_3f_3$, $f_1=e^{\theta_1}+e^{\theta_2}+e^{\theta_3}$, $\theta_i=k_ix+\omega_it+l_iy+s_iz+\delta_i$, $i=1,2,3$ into the Hirota bilinear form (\ref{Dx^m1D_y^m2D_t^m3D_z^m4}). We obtain the dispersion relation as
\begin{equation}\label{dispersionDx^m1D_y^m2D_t^m3D_z^m4}
k_i^{m_1}l_i^{m_2}\omega_i^{m_3}s_i^{m_4}=0,\quad i=1,2,3.
\end{equation}
For having nondegenerate phase shifts we take, for example, $k_1=l_2=\omega_3=0$. Then the three-soliton condition (\ref{3SC}) becomes
\begin{equation}\label{4lu3SC}
\phi [(-1)^{m_4}a^{m_4}+(-1)^{m_2+m_3}b^{m_4}+(-1)^{m_1+m_3}c^{m_4}+(-1)^{m_1+m_2}d^{m_4}]=0,
\end{equation}
where
\begin{align}
&\phi=(k_2^2-k_3^2)^{m_1}(l_3^2-l_1^2)^{m_2}(\omega_1^2-\omega_2^2)^{m_3}(k_2k_3)^{m_1}(l_1l_3)^{m_2}(\omega_1\omega_2)^{m_3},\\
&a=s_2s_3^3+s_2^3s_1-s_2^3s_3+s_1^3s_3-s_1s_3^3-s_1^3s_2,\\
&b=-s_2s_3^3+s_2^3s_1+s_2^3s_3+s_1^3s_3-s_1s_3^3-s_1^3s_2,\\
&c=s_2s_3^3+s_2^3s_1-s_2^3s_3-s_1^3s_3+s_1s_3^3-s_1^3s_2,\\
&d=s_2s_3^3-s_2^3s_1-s_2^3s_3+s_1^3s_3-s_1s_3^3+s_1^3s_2.
\end{align}

Let $m_4=1$. Then for $m_2+m_3$ and $m_1+m_3$ even, i.e., $m_1, m_2, m_3$ odd, the equation (\ref{4lu3SC}) is satisfied directly. Hence, for $m_4=1$ and $m_1, m_2, m_3$ odd,
the equation (\ref{Dx^m1D_y^m2D_t^m3D_z^m4}) has three-soliton solution.

Let $m_4\geq 2$. Consider the case when $m_4$ is even, say $m_4=2n$, $n\in \mathbb{N}$. Take two different sets for the particular values of $s_i$, $i=1, 2, 3$.
For $(s_1,s_2,s_3)=(1,1,2)$ we have $a=d=0$ and $b=c$ so the condition (\ref{4lu3SC}) turns to be
\begin{equation}
144^n\phi[(-1)^{m_2+m_3}+(-1)^{m_1+m_3}]=0,
\end{equation}
which is satisfied if $(-1)^{m_1+m_2}=-1$. For $(s_1,s_2,s_3)=(1,2,3)$, we have $d=0$ and the condition (\ref{4lu3SC}) becomes
\begin{equation}
144^n+(-1)^{m_2+m_3}2304^n+(-1)^{m_1+m_3}3600^n=0,
\end{equation}
which is not satisfied when $(-1)^{m_1+m_2}=-1$ for any $n\in \mathbb{N}$.

Now consider the case when $m_4\geq 2$ is odd, say $m_4=2n+1$, $n\in \mathbb{N}$. Similar with the even case, take particular values of $(s_1,s_2,s_3)$. For $(s_1,s_2,s_3)=(1,1,2)$ we have $a=d=0$, $b=c$, and therefore the condition (\ref{4lu3SC}) becomes
\begin{equation}\label{odd1}
12(144^n)\phi[(-1)^{m_1+m_3}-(-1)^{m_2+m_3}]=0,
\end{equation}
which holds when $(-1)^{m_1+m_2}=1$. For $(s_1,s_2,s_3)=(1,2,3)$, we have $d=0$. The condition (\ref{4lu3SC}) gives
\begin{equation}\label{odd2}
-12(144^n)-48(-1)^{m_2+m_3}2304^n+60(-1)^{m_1+m_3}3600^n=0,
\end{equation}
which is not satisfied when $(-1)^{m_1+m_2}=1$ for any nonzero $n\in \mathbb{N}$. For $n=0$, corresponding to $m_4=1$, if $m_1, m_2, m_3$ are odd then both
(\ref{odd1}) and (\ref{odd2}) are satisfied.

Hence the equation (\ref{Dx^m1D_y^m2D_t^m3D_z^m4}) has three-soliton solution if only if one of $m_i=1$, and all the other $m_i$'s are odd for $i=1,2,3,4$.

\noindent \textbf{Remark 4.} The equation (\ref{Dx^m1D_y^m2D_t^m3D_z^m4}) does not possess a genuine four-soliton solution.
We have showed that to have three-soliton solution one of $m_i=1$, say $m_4=1$, and $m_1, m_2, m_3$ are all odd.
We check four-soliton solution condition (\ref{4SC}) under this case with the dispersion relation, e.g. $s_1=0, l_2=0, \omega_3=0, k_4=0$. The four-soliton condition must be satisfied directly by any choice of parameters. Hence let us consider a particular case: $(k_1,k_2,k_3)=(1,2,3)$,  $(l_1,l_3,l_4)=(1,2,3)$, $(\omega_1,\omega_2,\omega_4)=(1,2,3)$, and $(s_2,s_3,s_4)=(1,2,3)$. In this case the condition (\ref{4SC}) becomes
\begin{equation}
(-1)^{m_3+1}72(6^{m_1+m_2+m_3})12^{m_1+m_2+m_3}[1+4^{m_1+m_2+m_3}+4(5^{m_1+m_3})]=0,
\end{equation}
which cannot be satisfied for any choice of $m_1, m_2, m_3$. Hence there exists no four-soliton solutions of the equation (\ref{Dx^m1D_y^m2D_t^m3D_z^m4}).

\section{Soliton solutions}

Since having Hirota bilinear form guarantees one- and two-soliton solutions,
here we only consider three-soliton solutions and, if any exists, four-soliton solutions as well.

\subsection{Three-soliton solutions of (\ref{Dx^mD_y^nD_t^p})}

Here let us consider the case for $m=3, p=1, n=2$, that is the equation (\ref{m=3p=1n=2}). We give the following particular example.

\noindent \textbf{Example 1.} Let us take $k_2=i, k_3=2i$, $l_1=2i$, $l_3=-i$, $\omega_1=8$, $\omega_2=-1$, and $\delta_i=0$ for $i=1, 2, 3$ in the three-soliton solution (\ref{three-solDx^mD_y^nD_t^p}). To sketch the graph of the solution
we consider the real-valued function $|u(x,y,t)|^2$ at $t=0$. We have $|u(x,y,0)|^2=\frac{W_1}{W_2}$, where
\begin{align}
&W_1=36[ 659486+5103\cos(y+2x)+7002\cos(2x-y)+148554\cos(2y)+92736\cos(x)+567\cos(2x-3y)\nonumber\\
&+7938\cos(x-2y)+10206\cos(2y+x)+99792\cos(x-y)+10206\cos(x-3y)+10206\cos(x-3y)\nonumber\\
&+71442\cos(x+y)]\\
&W_2=1550972+378216\cos(2x-y)+1323\cos(3x-3y)+220752\cos(x)+37422\cos(2x-3y)\nonumber\\
&+47628\cos(x-2y)+61236\cos(x+2y)+449064\cos(x-y)+45927\cos(x-3y)+321489\cos(x+y)\nonumber\\
&+16632\cos(3x-y)+15309\cos(y+3x)+336798\cos(y+2x)+403704\cos(2y).
\end{align}
The 3D and contour plot graphs of the solution $|u(x,y,0)|^2$ are given in the following Figure 1.
\begin{center}
\begin{figure}[h!]
\centering
\subfloat[]{\includegraphics[width=0.33\textwidth]{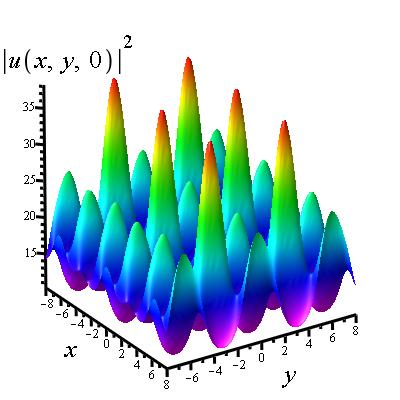}}\hspace{2cm}
\subfloat[]{\includegraphics[width=0.33\textwidth]{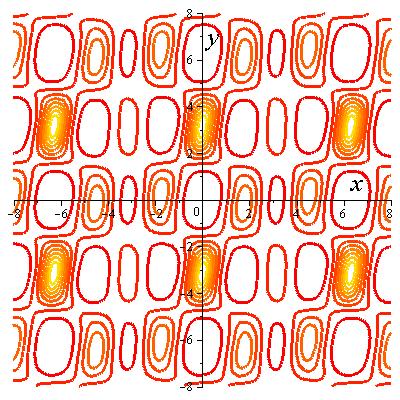}}
\caption{Breather-type wave solution of the equation  (\ref{m=3p=1n=2}) at $t=0$ (a) 3D graph, (b) contour plot.}
\end{figure}
\end{center}
\squeezeup

\subsection{Three- and four-soliton solutions of (\ref{bilinearProb2})}

Here, we consider two cases of (\ref{bilinearProb2}) that are (\ref{exk=0}) for $k=0$ and (\ref{exk=1}) for $k=1$. Let us first give a particular example for (\ref{exk=0}).

\noindent \textbf{Example 2.} Take $\alpha_1=1$, $\alpha_2=5$, $k_1=1, k_2=-2, k_3=3$, $l_1=2, l_2=l_3=1$, and $\delta_i=0$, $i=1, 2, 3$ in (\ref{three-solDx^mD_y^nD_t^p}) with (\ref{Prob2A_ij}). Then three-soliton solution of the equation (\ref{exk=0}) becomes
{\small \begin{equation}\label{threesolex1}\displaystyle
u(x,y,t)=\frac{2(e^{x-11t+2y}-2e^{-2x+3t+y}+3e^{3x-32t+y}-9e^{-x-8t+3y}+e^{4x-43t+3y}+25e^{x-29t+2y}+\frac{225}{2}e^{2x-40t+4y}       )}{1+e^{x-11t+2y}+e^{-2x+3t+y}+e^{3x-32t+y}+9e^{-x-8t+3y}+\frac{1}{4}e^{4x-43t+3y}+25e^{x-29t+2y}+\frac{225}{4}e^{2x-40t+4y}}.
\end{equation}}
The 3D and contour plot graphs of the above solution at $t=0$ are given in the following Figure 2.
\begin{center}
\begin{figure}[h!]
\centering
\subfloat[]{\includegraphics[width=0.33\textwidth]{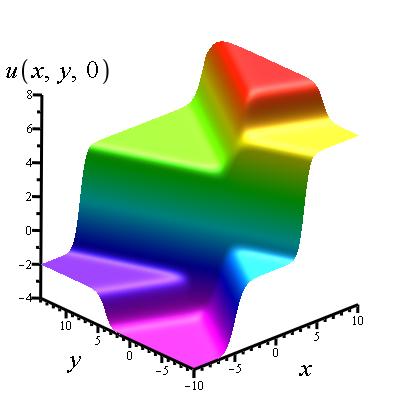}}\hspace{2cm}
\subfloat[]{\includegraphics[width=0.33\textwidth]{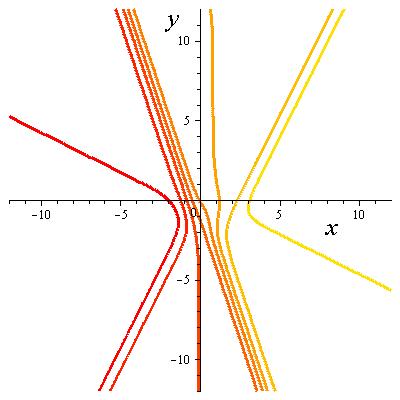}}
\caption{Three-soliton solution (\ref{threesolex1}) of the equation (\ref{exk=0}) at $t=0$ (a) 3D graph, (b) contour plot.}
\end{figure}
\end{center}
\squeezeup

\noindent Now, we give an example for the equation (\ref{exk=1}) that is for $k=1$.

\noindent \textbf{Example 3.}  Take $\alpha_1=-2$, $k_1=1, k_2=2, k_3=3$, and $\delta_i=0$, $i=1, 2, 3$ in (\ref{three-solDx^mD_y^nD_t^p}) with (\ref{Prob2A_ij}). Then three-soliton solution of the equation (\ref{exk=1}) turns to be
{\small \begin{equation}\label{threesolex2}\displaystyle
u(x,t)=\frac{2(e^{x+\frac{1}{2}t}+2e^{2x+4t}+3e^{3x+\frac{27}{2}t}+\frac{1}{27}e^{3x+\frac{9}{2}t}+\frac{1}{4}e^{4x+14t}+\frac{1}{125}e^{5x+\frac{35}{2}t}+\frac{1}{135000}e^{6x+18t}       )}{1+e^{x+\frac{1}{2}t}+e^{2x+4t}+e^{3x+\frac{27}{2}t}+\frac{1}{81}e^{3x+\frac{9}{2}t}+\frac{1}{16}e^{4x+14t}+\frac{1}{625}e^{5x+\frac{35}{2}t}+\frac{1}{810000}e^{6x+18t}}.
\end{equation}}
The 3D and contour plot graphs of the above solution are given in the following Figure 3.
\begin{center}
\begin{figure}[h!]
\centering
\subfloat[]{\includegraphics[width=0.33\textwidth]{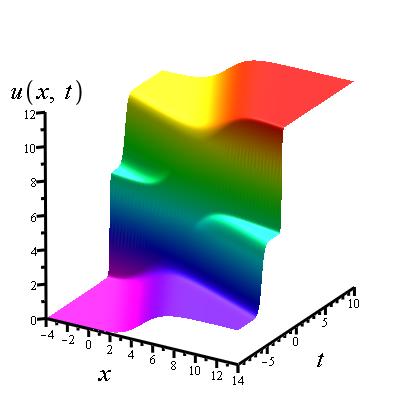}}\hspace{2cm}
\subfloat[]{\includegraphics[width=0.33\textwidth]{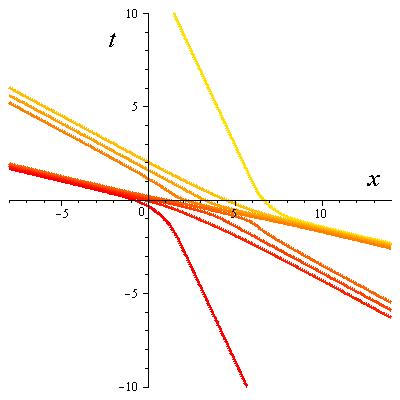}}
\caption{Three-soliton solution (\ref{threesolex2}) of the equation (\ref{exk=1}) (a) 3D graph, (b) contour plot.}
\end{figure}
\end{center}
\squeezeup

\noindent  In Section 2.1, we showed that the equation (\ref{bilinearProb2}) has four-soliton solution for $k=0$ but not for $k=1$. In the following example, we present
an example of four-soliton solution for the equation (\ref{exk=0}).

\noindent \textbf{Example 4.} Pick $\alpha_1=1$, $\alpha_2=5$, $k_1=1, k_2=-2, k_3=3$, $k_4=4$, $l_1=2, l_2=l_3=1$, $l_4=-2$, and $\delta_i=0$, $i=1, 2, 3, 4$ in (\ref{Prob24SS}). Then four-soliton solution of the equation (\ref{exk=0}) becomes
\begin{equation}\label{foursolex}\displaystyle
u(x,y,t)=2\frac{A(x,y,t)}{B(x,y,t)},
\end{equation}
where
{\small \begin{align}\displaystyle
&A(x,y,t)=e^{x-11t+2y}-2e^{-2x+3t+y}+3e^{3x-32t+y}+4e^{4x-54t-2y}-9e^{-x-8t+3y}+e^{4x-43t+3y}+\frac{9}{5}e^{5x-65t}\nonumber
\\&+25e^{x-29t+2y}+18e^{2x-51t-y}+\frac{1}{7}e^{7x-86t-y}
+\frac{2187}{98}e^{6x-94t+2y},\\
&B(x,y,t)=1+e^{x-11t+2y}+e^{-2x+3t+y}+e^{3x-32t+y}+e^{4x-54t-2y}+9e^{-x-8t+3y}+\frac{1}{4}e^{4x-43t+3y}+\frac{9}{25}e^{5x-65t}\nonumber
\\&+25e^{x-29t+2y}+9e^{2x-51t-y}+\frac{1}{49}e^{7x-86t-y}
+\frac{729}{196}e^{6x-94t+2y}.
\end{align}}
The 3D and contour plot graphs of the above solution at $t=0$ are given in the following Figure 4.
\begin{center}
\begin{figure}[h!]
\centering
\subfloat[]{\includegraphics[width=0.33\textwidth]{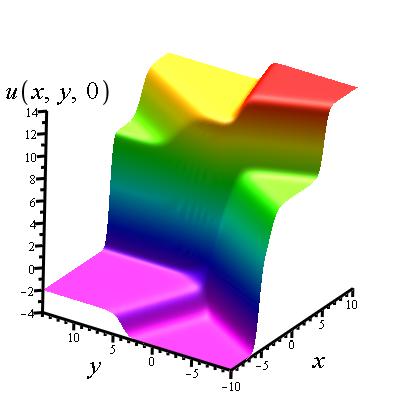}}\hspace{2cm}
\subfloat[]{\includegraphics[width=0.33\textwidth]{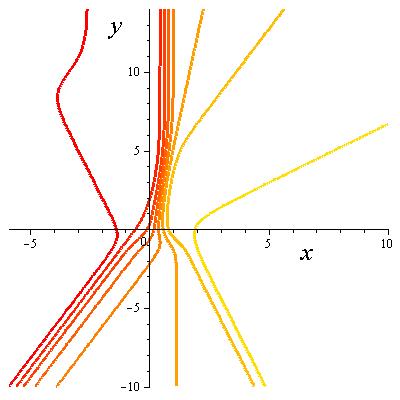}}
\caption{Four-soliton solution of the equation (\ref{exk=0}) at $t=0$ (a) 3D graph, (b) contour plot.}
\end{figure}
\end{center}
\squeezeup

\subsection{Three-soliton solution of the equation (\ref{bilinearDxDyDtDz})}

Let us consider the case when $m_i=1$, $i=1, 2, 3, 4$ in (\ref{Dx^m1D_y^m2D_t^m3D_z^m4}) that is the equation (\ref{bilinearDxDyDtDz}). Then we have the dispersion relation as
\begin{equation}
k_il_i\omega_is_i=0,\quad i=1, 2, 3.
\end{equation}
Hence, for instance, take $s_1=l_2=\omega_3=0$ to have nondegenerate phase shifts. Therefore we obtain three-soliton solution of the equation (\ref{eqnDxDyDtDz}) in the form (\ref{three-solDx^mD_y^nD_t^p}), where
\begin{align}
A_{12}=\frac{(k_1-k_2)(\omega_1-\omega_2)}{(k_1+k_2)(\omega_1+\omega_2)},\, A_{13}=\frac{(k_1-k_3)(l_1-l_3)}{(k_1+k_3)(l_1+l_3)}, \, A_{23}=\frac{(k_2-k_3)(s_2-s_3)}{(k_2+k_3)(s_2+s_3)}.
\end{align}
Consider now a particular example.

\noindent \textbf{Example 5.} Pick $k_1=4, k_2=2, k_3=1$, $\omega_1=2, \omega_2=-1$, $l_1=3, l_3=-1$, $s_2=2, s_3=-1$, $\delta_1=1, \delta_2=-1, \delta_3=2$, and $s_1=l_2=\omega_3=0$. We get three-soliton solution of (\ref{eqnDxDyDtDz}) as
\begin{equation}\displaystyle
u=\frac{V_1}{V_2},
\end{equation}
where
\begin{align}
V_1=&6[60e^{4x+2t+3y+1}+20e^{2x-t-2z-1}+15e^{x-y-z+2}+90e^{6x+t+3y-2z}+90e^{5x+2t+2y-z+3}\nonumber\\
&+5e^{3x-t-y-3z+1}+14e^{7x+t+2y-3z+2}],\nonumber\\
V_2=&45+5e^{3x-t-y-3z+1}+54e^{5x+2t+2y-z+3}+45e^{6x+t+3y-2z}+45e^{4x+2t+3y+1}+45e^{2x-t-2z-1}\nonumber\\
&+45e^{x-y-z+2}+6e^{7x+t+2y-3z+2}.
\end{align}
The 3D and contour graphs of the above solution at $y=z=0$ are given in Figure 5.
\begin{center}
\begin{figure}[h!]
\centering
\subfloat[]{\includegraphics[width=0.33\textwidth]{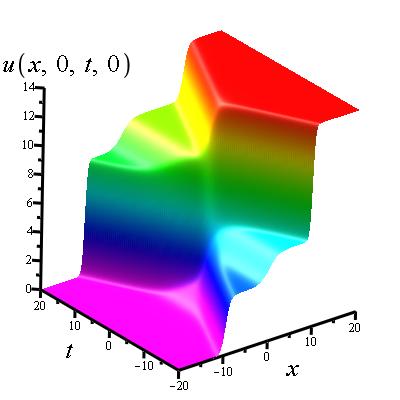}}\hspace{2cm}
\subfloat[]{\includegraphics[width=0.33\textwidth]{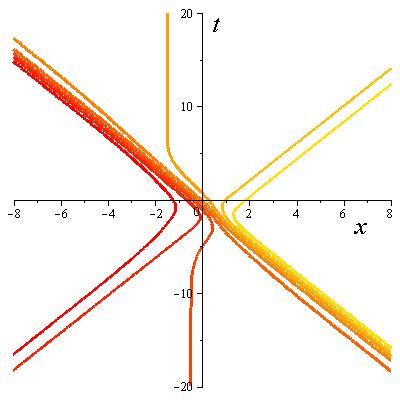}}
\caption{Three-soliton solution of (\ref{bilinearDxDyDtDz}) for the parameters $k_1=4, k_2=2, k_3=1$, $\omega_1=2, \omega_2=-1$, $l_1=3, l_3=-1$, $s_2=2, s_3=-1$, $\delta_1=1, \delta_2=-1, \delta_3=2$, and $s_1=l_2=\omega_3=0$  at $y=z=0$ (a) 3D graph, (b) contour plot.}
\end{figure}
\end{center}
\squeezeup

\section{Lump solutions }

\subsection{A. Lump solutions with two functions}

Let
\begin{equation}\label{lump2func}
f=\beta_0+p^2+q^2,
\end{equation}
where $p=\beta_1 x+\beta_2 y+\beta_3t+\beta_4$ and $q=\beta_5 x+\beta_6 y+\beta_7 t+\beta_8$ for some constants
$\beta_i$, $i=0, 1,\ldots, 8$.

\subsubsection{Lump solutions of the equation (\ref{Dx^mD_y^nD_t^p})}

Here consider the equation (\ref{m=3p=1n=2}) that is the case for $m=3, p=1, n=2$. The bilinear form of the equation (\ref{m=3p=1n=2}) is
\begin{equation}
D_x^3D_y^2D_t\{f\cdot f\}=\sum_{i=0}^3\sum_{j=0}^2\sum_{k=0}^1(-1)^{i+j+k}\binom{3}{i}\binom{2}{j}\binom{1}{k}f_{(3-i)x,(2-j)y,(1-k)t}f_{ix,jy,kt}=0,
\end{equation}
corresponding to
\begin{align}\label{bilinearformm=3p=1n=2}
&ff_{tyyxxx}-3f_xf_{tyyxx}+3f_{xx}f_{tyyx}-f_{xxx}f_{tyy}+6f_{xy}f_{tyxx}+f_{yy}f_{txxx}-f_tf_{yyxxx}-6f_{yxx}f_{xyt}\nonumber\\
&-2f_yf_{tyxxx}-3f_{yyx}f_{txx}+3f_{tx}f_{yyxx}+2f_{ty}f_{yxxx}=0.
\end{align}
Since each term involves derivatives of the function $f(x,y,t)$ at least order three and $f(x,y,t)$ is a polynomial of degree two, all of them vanishes directly.
There are no conditions on the parameters $\beta_i$, $i=0, 1,\ldots, 8$. Let us give the following example.

\noindent \textbf{Example 6.} Take the parameters as $\beta_0=2$, $\beta_1=1$, $\beta_2=-3$, $\beta_3=2$, $\beta_4=0$, $\beta_5=-2$, $\beta_6=2$, $\beta_7=4$, and
$\beta_8=5$. Hence lump solution of the equation (\ref{Dx^mD_y^nD_t^p}) for $m=3, p=1, n=2$ becomes
\begin{equation}\displaystyle
u(x,y,t)=\frac{4(5x-7y-6t-10)}{2+(x-3y+2t)^2+(-2x+2y+4t+5)^2}.
\end{equation}
The 3D and contour plot graphs of the above solution are given in the following Figure 6.
\begin{center}
\begin{figure}[h!]
\centering
\subfloat[]{\includegraphics[width=0.33\textwidth]{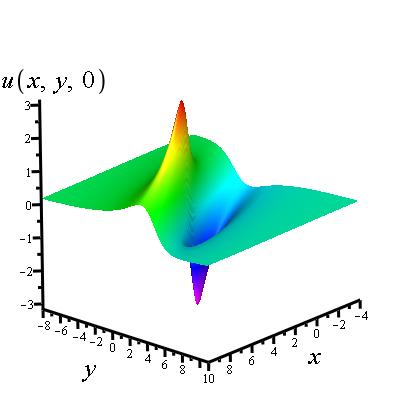}}\hspace{2cm}
\subfloat[]{\includegraphics[width=0.33\textwidth]{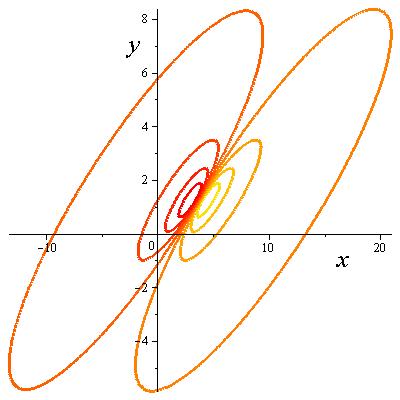}}
\caption{Lump solution of the equation (\ref{Dx^mD_y^nD_t^p}) for $m=3, p=1, n=2$ at $t=0$ (a) 3D graph, (b) contour plot.}
\end{figure}
\end{center}
\squeezeup

\subsubsection{Lump solutions of the equation (\ref{bilinearProb2})}

In this part we will only consider lump solutions for the cases $k=0$ and $k=1$ of the equation (\ref{bilinearProb2}).

\noindent \textbf{A.I. For k=0.}

Assume that $\alpha_1, \alpha_2$ are nonzero. Inserting (\ref{lump2func}) into $u=2(\ln f)_x$ for the equation (\ref{exk=0}) yields the following conditions on the parameters:
\begin{align}
&1)\, \alpha_1\beta_5^2\beta_7+\alpha_2\beta_5^2\beta_6-2\alpha_1\beta_1\beta_3\beta_5-\alpha_2\beta_1^2\beta_6-2\alpha_2\beta_1\beta_2\beta_5-\alpha_1\beta_1^2\beta_7=0,\\
&2)\, \alpha_1\beta_5^2\beta_3+\alpha_2\beta_5^2\beta_2+2\alpha_1\beta_1\beta_5\beta_7-\alpha_2\beta_1^2\beta_2+2\alpha_2\beta_1\beta_5\beta_6-\alpha_1\beta_1^2\beta_3=0,\\
&3)\, 3\alpha_1\beta_0\beta_1^2\beta_3+24\beta_1^3\beta_5^2+12\beta_1\beta_5^4+2\alpha_1\beta_0\beta_1\beta_5\beta_7+2\alpha_2\beta_0\beta_1\beta_5\beta_6
+3\alpha_2\beta_0\beta_1^2\beta_2+\alpha_1\beta_0\beta_3\beta_5^2+12\beta_1^5\nonumber\\
&+\alpha_2\beta_0\beta_2\beta_5^2=0,\\
&4)\, 24\beta_1^2\beta_5^3+3\alpha_2\beta_0\beta_5^2\beta_6+12\beta_1^4\beta_5+\alpha_1\beta_0\beta_1^2\beta_7+2\alpha_1\beta_0\beta_1\beta_3\beta_5
+2\alpha_1\beta_0\beta_1\beta_2\beta_5+12\beta_5^5+\alpha_2\beta_0\beta_1^2\beta_6\nonumber\\
&+3\alpha_1\beta_0\beta_5^2\beta_7=0.
\end{align}
The above conditions are satisfied under the following possibilities:
\begin{align}\displaystyle
&(i)\, \beta_1=i\beta_5,\, \beta_2=-\frac{\alpha_1\beta_3}{\alpha_2},\, \beta_6=-\frac{\alpha_1\beta_7}{\alpha_2}; \beta_0,\, \beta_3,\, \beta_4,\, \beta_5,\, \beta_7,\, \beta_8\, \mathrm{free}.\\
&(ii)\, \beta_0=0,\, \beta_1=i\beta_5,\, \beta_2=-\frac{i\alpha_1\beta_7+i\alpha_2\beta_6+\alpha_1\beta_3}{\alpha_2};  \beta_3,\, \beta_4,\, \beta_5,\, \beta_6,\, \beta_7,\, \beta_8\, \mathrm{free}.\\
&(iii)\, \beta_0=0,\, \beta_1=-i\beta_5,\, \beta_2=-\frac{-i\alpha_1\beta_7-i\alpha_2\beta_6+\alpha_1\beta_3}{\alpha_2};  \beta_3,\, \beta_4,\, \beta_5,\, \beta_6,\, \beta_7,\, \beta_8\, \mathrm{free}.\\
&(iv)\, \beta_1=\beta_5=0; \beta_0,\, \beta_2,\, \beta_3,\, \beta_4,\, \beta_5,\, \beta_6,\, \beta_7,\, \beta_8\, \mathrm{free}.
\end{align}
Consider the constraints $(i)$. Then we obtain the lump solution of the equation (\ref{exk=0}) as
\begin{equation}\label{lumpk=0}\displaystyle
u(x,y,t)=\frac{4\beta_5[i(i\beta_5 x-\frac{\alpha_1\beta_3}{\alpha_2}y+\beta_3t+\beta_4)+(\beta_5x-\frac{\alpha_1\beta_7}{\alpha_2}y+\beta_7t+\beta_8)   ]}
{[\beta_0+(i\beta_5 x-\frac{\alpha_1\beta_3}{\alpha_2}y+\beta_3t+\beta_4)^2+(\beta_5x-\frac{\alpha_1\beta_7}{\alpha_2}y+\beta_7t+\beta_8)^2]}.
\end{equation}

\noindent \textbf{Example 7.} Take the solution parameters as $\alpha_1=1$, $\alpha_2=5$, $\beta_0=1$, $\beta_3=2$, $\beta_4=1$, $\beta_5=2$, $\beta_7=4$, and $\beta_8=2$.
We obtain the following complex-valued solution
\begin{equation}\displaystyle
u(x,y,t)=\frac{4(-2iy+10it+10+5i-4y+20t)}{15+10ix-4ixy+20ixt+2y^2-20yt-10y+50t^2+50t-8xy+40xt+20x}.
\end{equation}
To sketch the graph of the solution we consider $|u(x,y,0)|^2$ which is
{\small \begin{equation}\displaystyle
|u(x,y,0)|^2=\frac{80(4y^2-20y+25)}{225+4y^4-40y^3+160y^2-300y+240y^2x+80x^2y^2-400x^2y-32y^3x-640xy+500x^2}.
\end{equation}}
The 3D and contour plot graphs of the above solution are given in the following Figure 7.
\begin{center}
\begin{figure}[h!]
\centering
\subfloat[]{\includegraphics[width=0.33\textwidth]{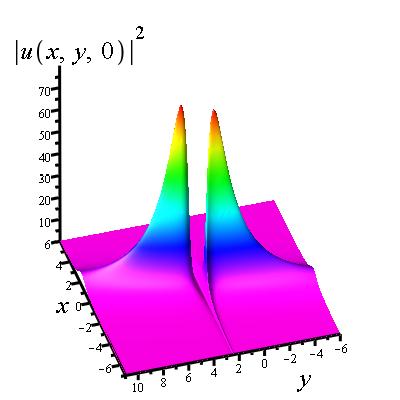}}\hspace{2cm}
\subfloat[]{\includegraphics[width=0.33\textwidth]{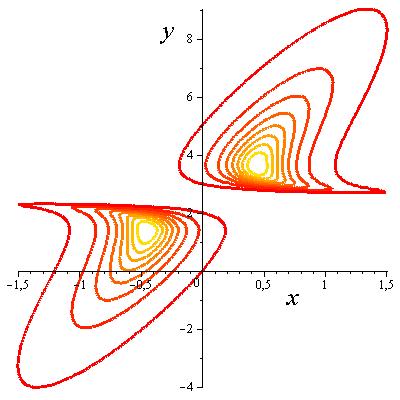}}
\caption{Lump solution of the equation (\ref{exk=0}) at $t=0$ (a) 3D graph, (b) contour plot.}
\end{figure}
\end{center}
\squeezeup

\noindent \textbf{A.II. For k=1.}

Note that by assuming $\alpha_2=0$ and $f=f(x,t)$ for this case in the form (\ref{lump2func}) we have $p=\beta_1 x+\beta_2t+\beta_3$ and $q=\beta_4 x+\beta_5 t+\beta_6$ for some constants $\beta_i$, $i=0, 1,\ldots, 6$. Using (\ref{lump2func}) in the equation (\ref{exk=1}) we obtain the following possibilities on the parameters:
\begin{align}\displaystyle
&(i)\, \beta_2=-\frac{\beta_4\beta_5}{\beta_1}; \beta_0,\, \beta_1,\, \beta_3,\, \beta_4,\, \beta_5,\, \beta_6\, \mathrm{free}.\\
&(ii)\, \beta_2=i\beta_5; \beta_0,\, \beta_1,\, \beta_3,\, \beta_4,\, \beta_5,\, \beta_6\, \mathrm{free}.\\
&(iii)\, \beta_2=-i\beta_5; \beta_0,\, \beta_1,\, \beta_3,\, \beta_4,\, \beta_5,\, \beta_6\, \mathrm{free}.\\
&(iv)\, \beta_1=\beta_5=0; \beta_0,\, \beta_2,\, \beta_3,\, \beta_4,\, \beta_6\, \mathrm{free}.
\end{align}
Consider the case $(i)$. Then we have the solution of the equation (\ref{exk=1}) as
\begin{equation}\label{lumpk=1}\displaystyle
u(x,t)=\frac{4[(\beta_1x-\frac{\beta_4\beta_5}{\beta_1}t+\beta_3)\beta_1+(\beta_4x+\beta_5t+\beta_6)\beta_4   ] }{\beta_0+(\beta_1x-\frac{\beta_4\beta_5}{\beta_1}t+\beta_3)^2+(\beta_4x+\beta_5t+\beta_6)^2}.
\end{equation}

\noindent \textbf{Example 8.} Pick the parameters as $\alpha_1=1$, $\beta_0=\beta_1=\beta_4=1$, $\beta_3=3$, $\beta_5=2$, and $\beta_6=4$.
Then the solution (\ref{lumpk=1}) becomes
\begin{equation}\displaystyle
u(x,t)=\frac{2(4x+14)}{1+(x-2t+3)^2+(x+2t+4)^2}.
\end{equation}
The 3D and contour plot graphs of the above solution are given in the following Figure 8.
\begin{center}
\begin{figure}[h!]
\centering
\subfloat[]{\includegraphics[width=0.33\textwidth]{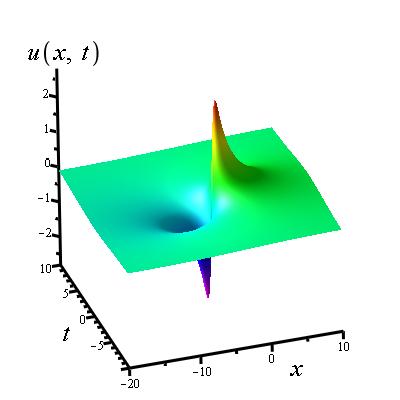}}\hspace{2cm}
\subfloat[]{\includegraphics[width=0.33\textwidth]{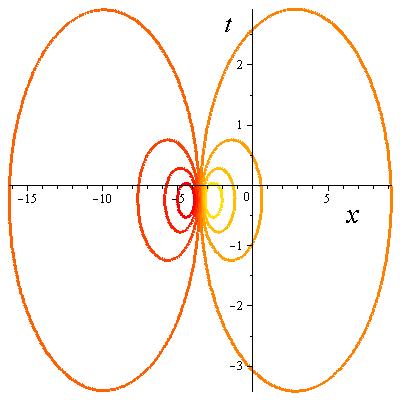}}
\caption{Lump solution of the equation (\ref{exk=1}). (a) 3D graph, (b) contour plot.}
\end{figure}
\end{center}
\squeezeup

\subsubsection{Lump solutions of the equation (\ref{bilinearDxDyDtDz})}

Since the Hirota bilinear form (\ref{bilinearDxDyDtDz}) represents a $(3+1)$-dimensional nonlinear partial differential equation we need to include the variable $z$, i.e., we take
$p=\beta_1 x+\beta_2 y+\beta_3t+\beta_4z+\beta_5$ and $q=\beta_6 x+\beta_7 y+\beta_8 t+\beta_9z+\beta_{10}$ for some constants
$\beta_i$, $i=0, 1,\ldots, 10$. The bilinear form (\ref{bilinearDxDyDtDz}) can be written as
\begin{align}\label{bilinearformxyzt}
&D_xD_yD_tD_z\{f\cdot f\}=\sum_{i,j,k,l\in\{0,1\}} (-1)^{i+j+k+l}f_{(1-i)x,(1-j)y,(1-k)t,(1-l)z}f_{ix,jy,kt,lz}\nonumber\\
&=2(ff_{xytz}-f_zf_{xyt}-f_tf_{xyz}-f_yf_{xzt}-f_xf_{ytz}+f_{xt}f_{yz}+f_{xy}f_{tz}+f_{xz}f_{yt})=0.
\end{align}
It is obvious that the terms that involve derivatives of order three or higher in the above equation vanish directly by (\ref{lump2func}) since the function
$f(x,y,t,z)$ is a polynomial of degree two. The terms $f_{xt}f_{yz}$, $f_{xy}f_{tz}$, and $f_{xz}f_{yt}$  yield the condition
\begin{equation}\label{beta_1}
\beta_1=\frac{\beta_6(\beta_3\beta_4\beta_7+\beta_2\beta_4\beta_8+3\beta_7\beta_8\beta_9+\beta_2\beta_3\beta_9 )}
{3\beta_2\beta_3\beta_4+\beta_2\beta_8\beta_9+\beta_3\beta_7\beta_9+\beta_4\beta_7\beta_8}.
\end{equation}
We can now present an example of lump solution for (\ref{bilinearDxDyDtDz}).

\noindent \textbf{Example 9.} Take $\beta_0=3i$, $\beta_2=-i$, $\beta_3=i$, $\beta_4=2i$, $\beta_5=0$, $\beta_6=2i$, $\beta_7=3i$, $\beta_8=-i$, $\beta_9=\beta_{10}=i$.
To sketch the graph of the solution we consider the real-valued solution $|u(x,y,t,z)|^2$ at $t=y=0$. Let us denote that function by $|g(x,z)|^2=|u(x,0,0,z)|^2$. We
have
\begin{equation}\displaystyle
|g(x,z)|^2=\frac{16[64+136xz+64z+16z^2+272x+289x^2]}{W(x,z)},
\end{equation}
where
\begin{align}
&W(x,z)=160+64z+128x+320xz+320z^3+544x^3+289x^4+400z^4+768xz^2+528zx^2\nonumber\\&+744x^2z^2+272x^3z+320xz^3+392x^2+224z^2.
\end{align}
The 3D and contour plot graphs of the above solution are given in the following Figure 9.
\begin{center}
\begin{figure}[h!]
\centering
\subfloat[]{\includegraphics[width=0.33\textwidth]{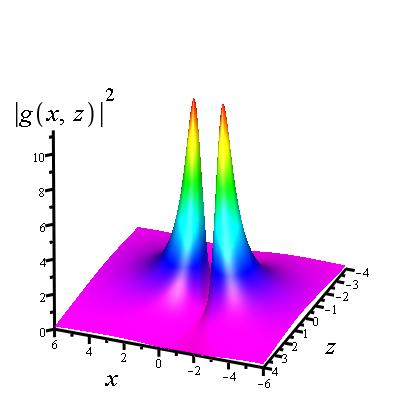}}\hspace{2cm}
\subfloat[]{\includegraphics[width=0.33\textwidth]{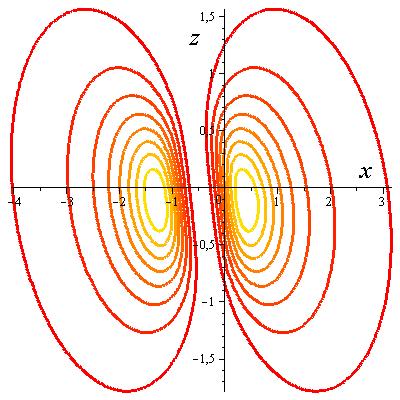}}
\caption{Lump solution of the equation (\ref{bilinearDxDyDtDz}) at $t=y=0$. (a) 3D graph, (b) contour plot.}
\end{figure}
\end{center}
\squeezeup

\section{B. Hybrid solutions}

\subsection{B(1). Hybrid solutions with $f=\beta_0+p^2+q$}
Let
\begin{equation}\label{hybrid1}
f=\beta_0+p^2+q,
\end{equation}
where $p=\beta_1 x+\beta_2 y+\beta_3t+\beta_4$ and $q=e^{\beta_5 x+\beta_6 y+\beta_7 t+\beta_8}$ for some constants
$\beta_i$, $i=0, 1,\ldots, 8$.

\subsubsection{B(1)-type hybrid solutions of (\ref{Dx^mD_y^nD_t^p})}

In this part we consider the particular case $m=3, p=1, n=2$ corresponding to the equation (\ref{m=3p=1n=2}) whose bilinear form is given by (\ref{bilinearformm=3p=1n=2}). Inserting (\ref{hybrid1}) into the
bilinear form (\ref{bilinearformm=3p=1n=2}) we obtain the following possibilities on the parameters:
\begin{align}
&i)\, \beta_6=\beta_7=0,\quad ii)\, \beta_3=\beta_7=0,\nonumber\\
&iii)\, \beta_2=\beta_6=0,\quad iv)\, \beta_5=0.
\end{align}
Let us take $\beta_5=0$ and give the following particular example.

\noindent \textbf{Example 10.} Choose $\beta_0=10$, $\beta_1=2$, $\beta_2=6$, $\beta_3=-4$, $\beta_4=1$, $\beta_6=-2$, $\beta_7=3$, $\beta_8=-1$. Hence hybrid solution of the equation (\ref{m=3p=1n=2}) becomes
\begin{equation}\displaystyle
u(x,y,t)=\frac{8(2x+6y-4t+1)}{10+(2x+6y-4t+1)^2+e^{-2y+3t-1}}.
\end{equation}
The 3D and contour plot graphs of the above solution at $t=0$ are given in the following Figure 10.
\begin{center}
\begin{figure}[h!]
\centering
\subfloat[]{\includegraphics[width=0.33\textwidth]{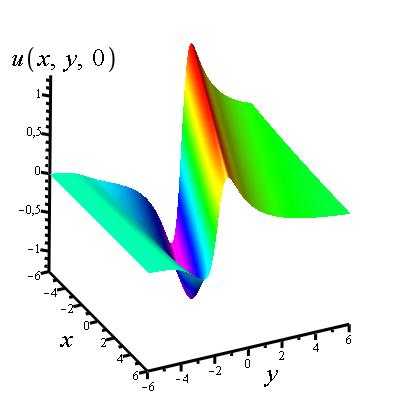}}\hspace{2cm}
\subfloat[]{\includegraphics[width=0.33\textwidth]{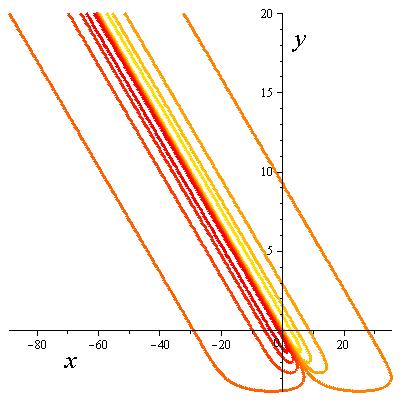}}
\caption{Hybrid lump solution of the equation (\ref{Dx^mD_y^nD_t^p}) for $m=3, p=1, n=2$ at $t=0$. (a) 3D graph, (b) contour plot.}
\end{figure}
\end{center}
\squeezeup

\subsubsection{B(1)-type hybrid solutions of (\ref{bilinearProb2})}

\noindent \textbf{B(1).I. For k=0.}

Assume that $\alpha_1, \alpha_2$ are nonzero. When we insert (\ref{hybrid1}) into (\ref{exk=0}) with $u=2(\ln f)_x$ we obtain the following constraints:
\begin{align}\displaystyle
&(i)\, \beta_1=0,\, \beta_2=-\frac{\alpha_1\beta_3}{\alpha_2},\, \beta_6=-\frac{\alpha_1\beta_7+\beta_5^3}{\alpha_2}; \beta_0,\, \beta_3,\, \beta_4,\, \beta_5,\, \beta_7,\, \beta_8\, \mathrm{free}.\\
&(ii)\, \beta_1=\beta_5=0; \beta_0,\, \beta_2,\, \beta_3,\, \beta_4,\, \beta_6,\, \beta_7,\, \beta_8\, \mathrm{free}.
\end{align}
The second case above gives trivial solution. Hence consider the constraints given in $(i)$.\\

\textbf{Example 11.} Take the solution parameters as $\alpha_1=1$, $\alpha_2=5$, $\beta_0=10$, $\beta_3=3$, $\beta_4=2$, $\beta_5=4$, $\beta_7=1$, and $\beta_8=2$.
Hence hybrid solution of the equation (\ref{exk=0}) becomes
\begin{equation}\displaystyle
u(x,y,t)=\frac{8e^{4x-13y+t+2}}{10+\Big(-\frac{3}{5}y+3t+2 \Big)^2+e^{4x-13y+t+2}}.
\end{equation}
The 3D and contour plot graphs of the above solution at $t=0$ are given in the following Figure 11.
\begin{center}
\begin{figure}[h!]
\centering
\subfloat[]{\includegraphics[width=0.33\textwidth]{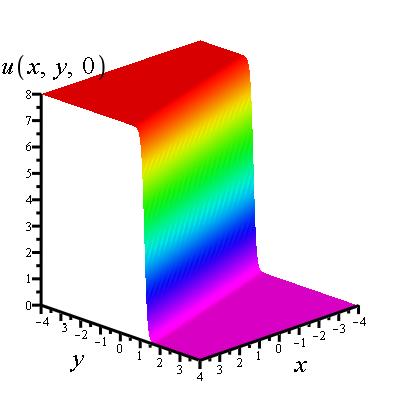}}\hspace{2cm}
\subfloat[]{\includegraphics[width=0.33\textwidth]{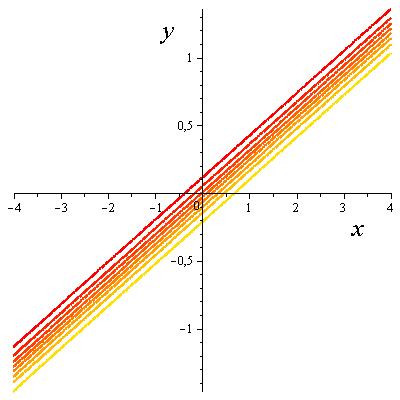}}
\caption{Kink-type wave solution of the equation (\ref{exk=0}) at $t=0$ (a) 3D graph, (b) contour plot.}
\end{figure}
\end{center}
\squeezeup

\noindent \textbf{B(1).II. For k=1.}

For the case $k=1$ we assume that $f=f(x,t)$ hence in (\ref{hybrid1}) we take $p=\beta_1 x+\beta_2t+\beta_3$ and $q=e^{\beta_4 x+\beta_5 t+\beta_6}$ for some constants
$\beta_i$, $i=0, 1,\ldots, 8$.\\

\textbf{Example 12.} Take the solution parameters as $\alpha_1=8$, $\beta_0=10i$, $\beta_2=i$, $\beta_3=3i$, $\beta_4=2i$, and $\beta_6=4i$.
Then the real-valued solution for the equation (\ref{exk=1}) is obtained as
\begin{equation}
|u(x,t)|^2=\frac{16}{182+108t+54t^2+12t^3+t^4-(2t^2+12t+18)\cos(2x+t+4)+20\sin(2x+t+4)}.
\end{equation}
The 3D and contour plot graphs of the above solution are given in the following Figure 12.
\begin{center}
\begin{figure}[h!]
\centering
\subfloat[]{\includegraphics[width=0.33\textwidth]{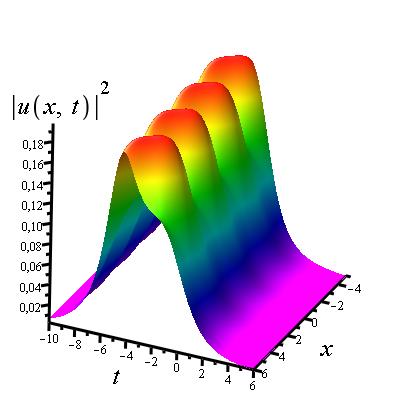}}\hspace{2cm}
\subfloat[]{\includegraphics[width=0.33\textwidth]{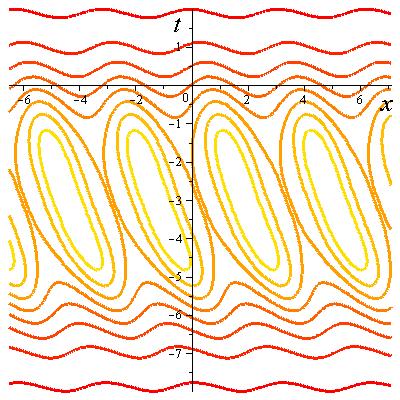}}
\caption{Periodic-type wave solution of the equation (\ref{exk=1}) (a) 3D graph, (b) contour plot.}
\end{figure}
\end{center}
\squeezeup

\subsubsection{B(1)-type hybrid solutions of (\ref{bilinearDxDyDtDz})}

For the equation (\ref{bilinearDxDyDtDz}) whose bilinear form given by (\ref{bilinearformxyzt}) we take
$p=\beta_1 x+\beta_2 y+\beta_3t+\beta_4z+\beta_5$ and $q=e^{\beta_6 x+\beta_7 y+\beta_8 t+\beta_9z+\beta_{10}}$ for some constants
$\beta_i$, $i=0, 1,\ldots, 10$. Inserting (\ref{hybrid1}) with these $p$ and $q$ functions yields the following possibilities on the parameters:
\begin{align}
&i)\, \beta_2=\beta_6=\beta_8=\beta_9=0,\quad ii)\, \beta_4=\beta_6=\beta_7=\beta_8=0, \nonumber\\
&iii)\, \beta_1=\beta_7=\beta_8=\beta_9=0,\quad iv)\, \beta_3=\beta_6=\beta_7=\beta_9=0,\nonumber\\
&v)\, \beta_1=\beta_6=0,\quad vi)\beta_3=\beta_8=0,\quad  vii)\beta_2=\beta_7=0,\quad viii)\beta_4=\beta_9=0.
\end{align}
Consider the case $i)$ where $\beta_2=\beta_6=\beta_8=\beta_9=0$. Let us give the following example.

\textbf{Example 13.} Choose the solution parameters as $\beta_0=10i$, $\beta_1=2i$ $\beta_3=6i$, $\beta_4=-4i$, $\beta_5=i$, $\beta_7=-2i$, and $\beta_{10}=3i$.
To sketch the graph of the solution we consider the real-valued solution $|u(x,y,t,z)|^2$ at $t=z=0$. Let us denote that function by $|g(x,y)|^2=|u(x,y,0,0)|^2$. We
have
\begin{equation}\displaystyle
|g(x,y)|^2=\frac{128x^2+128x+32}{51+12x^2+4x+8x^4+16x^3-10\sin(2y-3)-(4x^2+4x+1)\cos(2y-3)}.
\end{equation}
The 3D and contour plot graphs of the above solution are given in the following Figure 13.
\begin{center}
\begin{figure}[h!]
\centering
\subfloat[]{\includegraphics[width=0.33\textwidth]{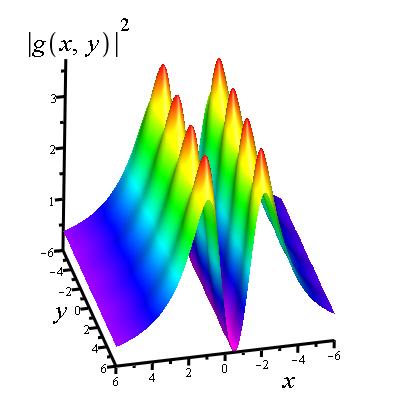}}\hspace{2cm}
\subfloat[]{\includegraphics[width=0.33\textwidth]{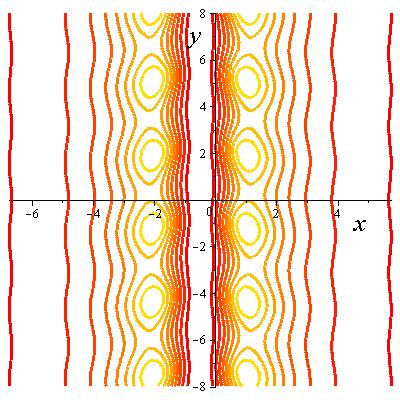}}
\caption{Periodic-type wave solution of the equation (\ref{bilinearDxDyDtDz}) at $t=z=0$ (a) 3D graph, (b) contour plot.}
\end{figure}
\end{center}
\squeezeup

\subsection{B(2). Hybrid solutions with $f=\beta_0+p^2+q+Ap^2q$}

Let
\begin{equation}\label{hybrid2}
f=\beta_0+p^2+q+Ap^2q,
\end{equation}
where $p=\beta_1 x+\beta_2 y+\beta_3t+\beta_4$ and $q=e^{\beta_5 x+\beta_6 y+\beta_7 t+\beta_8}$ for some constants
$\beta_i$, $i=0,1,\ldots, 8$.

\subsubsection{B(2)-type hybrid solutions of (\ref{Dx^mD_y^nD_t^p})}

Here, we again consider the particular case $m=3, p=1, n=2$ that is the equation (\ref{m=3p=1n=2}) with bilinear form (\ref{bilinearformm=3p=1n=2}). We insert (\ref{hybrid2}) into (\ref{bilinearformm=3p=1n=2}) and get the following possibilities on the solution parameters:
\begin{align}
&i)\, \beta_1=\beta_6=\beta_7=0,\quad ii)\, \beta_3=\beta_7=0,\quad iii)\, \beta_1=\beta_5=0,\nonumber\\
&iv)\, \beta_5=\beta_6,\quad v)\, \beta_0=\frac{1}{A}, \beta_1=\beta_2=\beta_7=0,\quad vi)\, \beta_3=-\frac{2\beta_7\beta_2}{\beta_6}, \beta_5=0.
\end{align}
Consider the case $vi)$ that is $\beta_3=-\frac{2\beta_7\beta_2}{\beta_6}, \beta_5=0$. We give the following example.

\textbf{Example 14.} Choose the solution parameters as $\beta_0=10$, $\beta_1=2$, $\beta_2=-1$, $\beta_4=1$, $\beta_6=-4$, $\beta_7=1$, $\beta_8=-1$, and $A=4$.
Hence we obtain the hybrid solution of the equation (\ref{m=3p=1n=2}) as
\begin{equation}\displaystyle
u(x,y,t)=\frac{2[8x-4y-2t+4+16(2x-y-\frac{1}{2}t+1)e^{-4y+t-1}]}{10+(2x-y-\frac{1}{2}t+1)^2+e^{-4y+t-1}+4(2x-y-\frac{1}{2}t+1)^2e^{-4y+t-1}}.
\end{equation}
The 3D and contour plot graphs of the above solution at $t=0$ are given in the following Figure 14.
\begin{center}
\begin{figure}[h!]
\centering
\subfloat[]{\includegraphics[width=0.33\textwidth]{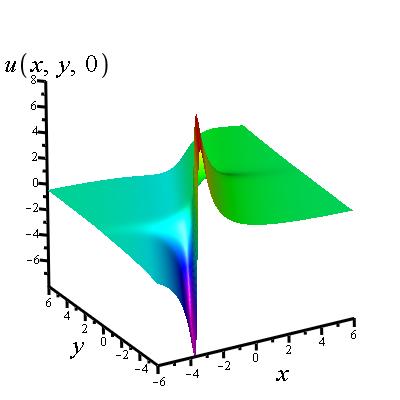}}\hspace{2cm}
\subfloat[]{\includegraphics[width=0.33\textwidth]{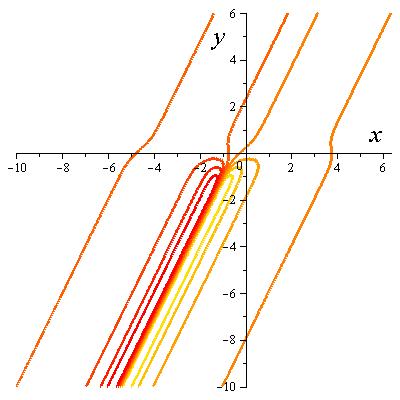}}
\caption{Hybrid lump solution of the equation (\ref{Dx^mD_y^nD_t^p}) for $m=3, p=1, n=2$ at $t=0$ (a) 3D graph, (b) contour plot.}
\end{figure}
\end{center}
\squeezeup

\subsubsection{B(2)-type hybrid solutions of (\ref{bilinearProb2})}

\noindent \textbf{B(2).I. For k=0.}

Assume that $\alpha_1, \alpha_2$ are nonzero. Insert (\ref{hybrid2}) into (\ref{exk=0}) with $u=2(\ln f)_x$ we obtain the following constraints:
\begin{align}\displaystyle
&(i)\, \beta_1=0,\, \beta_2=-\frac{\alpha_1\beta_3}{\alpha_2},\, \beta_6=-\frac{\alpha_1\beta_7+\beta_5^3}{\alpha_2}; A,\, \beta_0,\, \beta_3,\, \beta_4,\, \beta_5,\, \beta_7,\, \beta_8\, \mathrm{free}.\\
&(ii)\, \beta_0=\frac{1}{A},\, \beta_1=0,\, \beta_6=-\frac{\alpha_1\beta_7+\beta_5^3}{\alpha_2}; A,\, \beta_2,\, \beta_3,\, \beta_4,\, \beta_5,\, \beta_7,\, \beta_8\, \mathrm{free}.\\
&(iii)\, \beta_1=\beta_5=0; \beta_0,\, \beta_2,\, \beta_3,\, \beta_4,\, \beta_6,\, \beta_7,\, \beta_8\, \mathrm{free}.
\end{align}
Consider the case given in $(i)$.

\textbf{Example 15.} Pick the parameters as $\alpha_1=1$, $\alpha_2=5$, $\beta_0=10i$, $\beta_3=3i$, $\beta_4=2$, $\beta_5=i$, $\beta_7=i$, $\beta_8=2$, and $A=2$. Thus we obtain a complex-valued solution. To draw the graph of the solution we present the real-valued hybrid solution of the equation (\ref{exk=0}) at $t=0$ as
\begin{equation}\displaystyle
|u(x,y,0)|^2=\frac{36e^4(5625+700y^2+36y^4)}{W_1(x,y)},
\end{equation}
where
\begin{align}
&W_1(x,y)=72500+1800y^2+45000e^2\cos(x)+112500e^2\sin(x)-9000e^2y^2\sin(x)+6750e^2y^2\cos(x)
\nonumber\\
&-3000e^2y\sin(x)-60000e^2y\cos(x)324e^2y^4\cos(x)+50625e^4-30000y+81y^4+324e^4y^4+6300e^4y^2.
\end{align}
The 3D and contour plot graphs of the solution $|u(x,y,0)|^2$ are given in the following Figure 15.
\begin{center}
\begin{figure}[h!]
\centering
\subfloat[]{\includegraphics[width=0.33\textwidth]{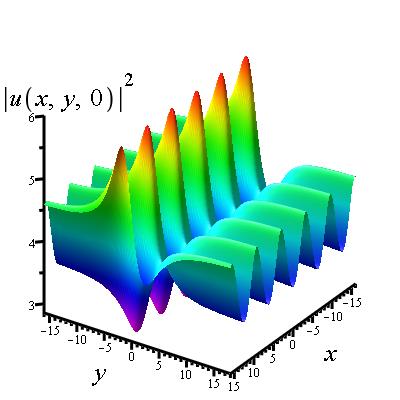}}\hspace{2cm}
\subfloat[]{\includegraphics[width=0.33\textwidth]{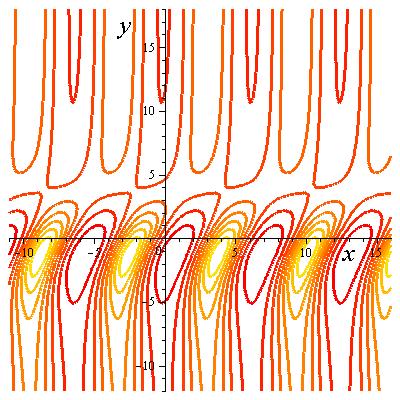}}
\caption{Periodic-type wave solution of the equation (\ref{exk=0}) at $t=0$ (a) 3D graph, (b) contour plot.}
\end{figure}
\end{center}
\squeezeup

\noindent \textbf{B(2).II. For k=1.}

For the case $k=1$ we assume that $f=f(x,t)$ hence in (\ref{hybrid2}) we take $p=\beta_1 x+\beta_2t+\beta_3$ and $q=e^{\beta_4 x+\beta_5 t+\beta_6}$ for some constants
$\beta_i$, $i=0, 1,\ldots, 8$.

\textbf{Example 16.} Pick $\alpha_1=1$, $\beta_0=10$, $\beta_2=-2$, $\beta_3=3$, $\beta_4=2$, $\beta_6=4$, and $A=2$. Therefore we obtain the following
kink-type wave soliton for the equation (\ref{exk=1}) as
\begin{equation}\displaystyle
u(x,t)=\frac{4(e^{2x-8t+4}+2(-2t+3)^2e^{2x-8t+4} ) }{10+(-2t+3)^2+e^{2x-8t+4}+2(-2t+3)^2e^{2x-8t+4}}
\end{equation}
The 3D and contour plot graphs of the solution are given in the following Figure 16.
\begin{center}
\begin{figure}[h!]
\centering
\subfloat[]{\includegraphics[width=0.33\textwidth]{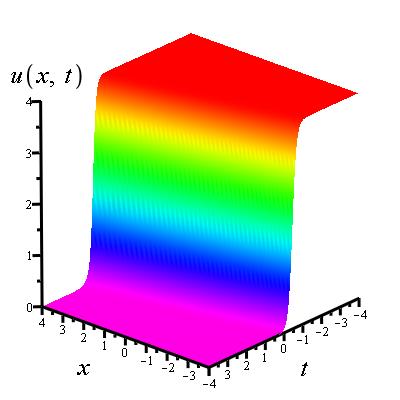}}\hspace{2cm}
\subfloat[]{\includegraphics[width=0.33\textwidth]{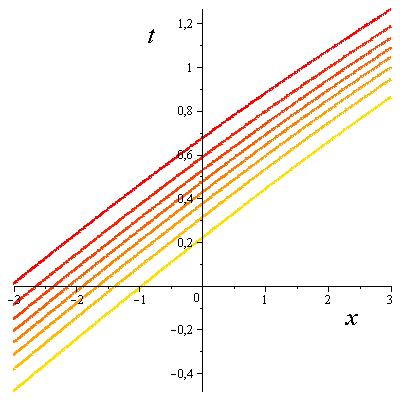}}
\caption{Kink-type wave solution of the equation (\ref{exk=1}) (a) 3D graph, (b) contour plot.}
\end{figure}
\end{center}
\squeezeup

\subsubsection{B(2)-type hybrid solutions of (\ref{bilinearDxDyDtDz})}

For the equation (\ref{bilinearDxDyDtDz}) with bilinear form (\ref{bilinearformxyzt}) we take
$p=\beta_1 x+\beta_2 y+\beta_3t+\beta_4z+\beta_5$ and $q=e^{\beta_6 x+\beta_7 y+\beta_8 t+\beta_9z+\beta_{10}}$ for some constants
$\beta_i$, $i=0, 1,\ldots, 10$. We insert (\ref{hybrid2}) into (\ref{bilinearformxyzt} with these $p$ and $q$ functions gives
the following possibilities on the parameters:
\begin{align}
&i)\,\beta_1=\beta_7=\beta_8=\beta_9=0,\quad ii)\,\beta_2=\beta_6=\beta_8=\beta_9=0,\quad iii)\,\beta_4=\beta_6=\beta_7=\beta_8=0,\nonumber\\
&iv)\,\beta_3=\beta_6=\beta_7=\beta_9=0,\quad v)\,\beta_3=\beta_6=\beta_8,\quad vi)\,\beta_3=\beta_7=\beta_8=0,\nonumber\\
&vii)\,\beta_4=\beta_6=\beta_9=0,\quad viii)\,\beta_4=\beta_7=\beta_9=0,\quad ix)\,\beta_2=\beta_7=0,\quad x)\,\beta_1=\beta_6=0,\nonumber\\
&xi)\,\beta_0=\frac{1}{A}, \beta_3=-\frac{\beta_4\beta_8}{\beta_9}, \beta_1=\beta_7=0,\quad xii)\,\beta_0=\frac{1}{A}, \beta_1=-\frac{\beta_3\beta_6}{\beta_8}, \beta_4=\beta_7=0,\nonumber\\
&xiii)\beta_0=\frac{1}{A}, \beta_2=-\frac{\beta_3\beta_7}{\beta_8}, \beta_4=\beta_6=0,\quad xiv)\beta_0=\frac{1}{A}, \beta_4=-\frac{\beta_2\beta_9}{\beta_7}, \beta_3=\beta_6=0,\nonumber\\
&xv)\beta_0=\frac{1}{A}, \beta_4=-\frac{\beta_1\beta_9}{\beta_6}, \beta_3=\beta_7=0,\quad xvi)\beta_0=\frac{1}{A}, \beta_4=-\frac{\beta_3\beta_9}{\beta_8}, \beta_2=\beta_6=0.
\end{align}
Pick the case $xiv)$, i.e., $\beta_0=\frac{1}{A}, \beta_4=-\frac{\beta_2\beta_9}{\beta_7}$, $\beta_3=\beta_6=0$. Consider the below example.

\textbf{Example 17.} Choose the solution parameters as $\beta_0=10i$, $\beta_1=2i$ $\beta_3=6i$, $\beta_4=-4i$, $\beta_5=i$, $\beta_7=-2i$, and $\beta_{10}=3i$.
have
\begin{equation}\displaystyle
u(x,y,t,z)=\frac{2[8x+24y-24z-16+4(2x+6y-6z-4)e^{y-2t+z+3}]}{1+(2x+6y-6z-4)^2+e^{y-2t+z+3}+(2x+6y-6z-4)^2e^{y-2t+z+3}}.
\end{equation}
The 3D and contour plot graphs of the above solution $t=z=0$ are given in the following Figure 17.
\begin{center}
\begin{figure}[h!]
\centering
\subfloat[]{\includegraphics[width=0.33\textwidth]{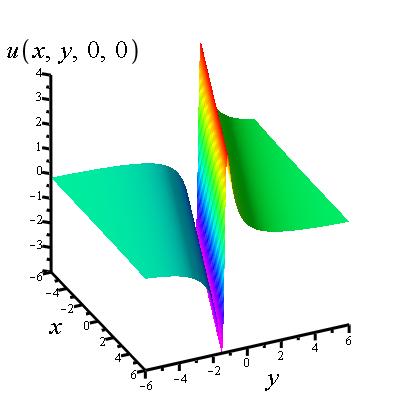}}\hspace{2cm}
\subfloat[]{\includegraphics[width=0.33\textwidth]{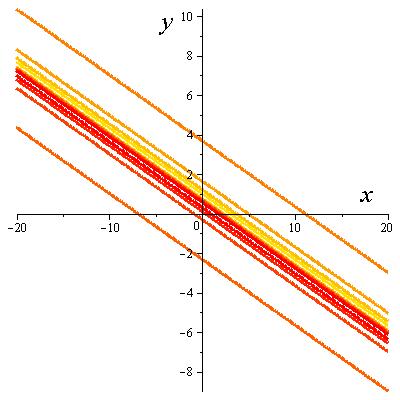}}
\caption{Hybrid lump-type wave solution of the equation (\ref{bilinearDxDyDtDz}) at $t=z=0$ (a) 3D graph, (b) contour plot.}
\end{figure}
\end{center}
\squeezeup

\section{Conclusion}
In this work we considered two $(2+1)$-dimensional Hirota bilinear forms; $D_x^mD_y^rD_t^n\{f\cdot f\}=0$ and $D_x(D_x^3+\alpha_1 D_t+\alpha_2 D_y)^{2k+1}\{f\cdot f\}=0$, and one $(3+1)$-dimensional  $D_x^{m_1}D_y^{m_2}D_t^{m_3}D_z^{m_4}\{f\cdot f\}=0$. We gave the proof for $P(D)=D_x^mD_y^rD_t^n$ that we have nonlinear partial differential
equations possessing three-soliton solution if only if two of nonzero $m,n,p$ are odd and the other one even. We also wrote these equations explicitly for $m+n+p=4$ and $m+n+p=6$ by the help of the list derived in Appendix covering all monomials up to sixth order operators $D_{x}$, $D_{y}$, and $D_{t}$. For the $(3+1)$-dimensional version of that form, i.e., $P(D)=D_x^{m_1}D_y^{m_2}D_t^{m_3}D_z^{m_4}$ we showed that these equations have three-soliton solutions if only if one of $m_i=1$, and all the other $m_i$'s are odd for $i=1,2,3,4$. We wrote the equation for $m_i=1$, $i=1, 2, 3, 4$ explicitly. We observed that these $(2+1)$-dimensional and $(3+1)$-dimensional Hirota bilinear forms do not have genuine four-soliton solutions.

In addition to them we presented a new class of equations given with $P(D)=D_x(D_x^3+\alpha_1 D_t+\alpha_2 D_y)^{2k+1}$ having three-soliton solution for any $k \in \mathbb{N}$. We gave the conjecture for these equations that the only equation having four-soliton solution is the one for $k=0$.

We also obtained three-soliton, lump, and hybrid solutions of particular cases of these equations; $D_x^3D_y^2D_t$, $D_xD_yD_tD_z$, $D_x(D_x^3+\alpha_1 D_t+\alpha_2 D_y)$, and $D_x(D_x^3+\alpha_1 D_t)^{3}$.

In our future work we will obtain other type of solutions such as complexitons, positons, negatons, and rogue waves especially for the new equations represented by
$D_x(D_x^3+\alpha_1 D_t+\alpha_2 D_y)^{2k+1}\{f\cdot f\}=0$. We plan to continue our study to construct new equations admitting three-soliton and even $N$-soliton solutions.

\section*{Appendix: Fourth and sixth order bilinear forms}

In this Appendix we give a list of monomials in Hirota bilinear forms and their associated nonlinear partial differential equations in $(2+1)$-dimensions. This list covers all monomials up to sixth order operators $D_{x}$, $D_{y}$, and $D_{t}$.
Under the transformation $u(x,y,t)=2(\ln(f(x,y,t)))_x$ we have the following equalities for the
monomials of binary products $D^{k}_{x}\,D^{2n-k}_{t}$, $D^{k}_{y}\,D^{2n-k}_{t}$, and $D^{k}_{x}\,D^{2n-k}_{y}$ for  $k=0,1,2, \cdots, 2n$, and $n=1,2,3$.\\

\noindent
{\bf I) Hirota bilinear forms with binary products}

\vspace{0.5cm}
\noindent
For $n=1$: All monomials are second order.
{\small\begin{align}
&\Big(\frac{D_x^2\{f\cdot f\}}{f^2}\Big)_x=u_{xx},~~\Big(\frac{D_x\,D_{t}\{f\cdot f\}}{f^2}\Big)_x=u_{xt},\\
&\Big(\frac{D_y^2\{f\cdot f\}}{f^2}\Big)_x=u_{yy},~~\Big(\frac{D_x\, D_{y}\{f\cdot f\}}{f^2}\Big)_x=u_{xy},\\
&\Big(\frac{D_t^2\{f\cdot f\}}{f^2}\Big)_x=u_{tt},~~\Big(\frac{D_y\,D_t\{f\cdot f\}}{f^2}\Big)_x=u_{yt}.
\end{align}}
For $n=2$:  All monomials are fourth order.
{\small\begin{align}
&\Big(\frac{D_x^3D_y\{f\cdot f\}}{f^2}\Big)_x=u_{xxxy}+3u_xu_{xy}+3u_yu_{xx},\\
&\Big(\frac{D_y^3D_x\{f\cdot f\}}{f^2}\Big)_x=u_{yyyx}+3u_yu_{yy}+3u_{xy} (D^{-1} u_{yy}),\\
&\Big(\frac{D_t^3D_x\{f\cdot f\}}{f^2}\Big)_x=u_{tttx}+3u_tu_{tt}+3u_{xt} (D^{-1} u_{tt}),\\
&\Big(\frac{D_t^3D_y\{f\cdot f\}}{f^2}\Big)_x=u_{ttty}+3u_{ty}(D^{-1}u_{tt})+3u_{tt}(D^{-1} u_{ty}),\\
&\Big(\frac{D_y^3D_t\{f\cdot f\}}{f^2}\Big)_x=u_{yyyt}+3u_{yt}(D^{-1} u_{yy})+3u_{yy}(D^{-1} u_{yt}),\\
&\Big(\frac{D_x^2D_t^2\{f\cdot f\}}{f^2}\Big)_x=u_{xxtt}+4u_tu_{xt}+u_xu_{tt}+u_{xx}(D^{-1}u_{tt}),\\
&\Big(\frac{D_x^2D_y^2\{f\cdot f\}}{f^2}\Big)_x=u_{xxyy}+4u_yu_{xy}+u_xu_{yy}+u_{xx}(D^{-1} u_{yy}),\\
&\Big(\frac{D_y^2D_t^2\{f\cdot f\}}{f^2}\Big)_x=u_{yytt}+u_{tt}(D^{-1} u_{yy})+u_{yy}(D^{-1} u_{tt})+4u_{yt}(D^{-1}u_{yt}),\\
&\Big(\frac{D_x^4\{f\cdot f\}}{f^2}\Big)_x=u_{xxxx}+6u_xu_{xx},\\
&\Big(\frac{D_y^4\{f\cdot f\}}{f^2}\Big)_x=u_{yyyy}+6u_{yy}(D^{-1}u_{yy}),\\
&\Big(\frac{D_t^4\{f\cdot f\}}{f^2}\Big)_x=u_{tttt}+6u_{tt}(D^{-1}u_{tt}).
\end{align}}

For $n=3$:  All monomials are sixth order.
{\small\begin{align}
&\Big(\frac{D_x^6\{f\cdot f\}}{f^2}\Big)_x=u_{6x}+15u_xu_{xxxx}+15u_{xx}u_{xxx}+45u_x^2u_{xx},\\
&\Big(\frac{D_y^6\{f\cdot f\}}{f^2}\Big)_x=u_{6y}+15u_{yyyy}(D^{-1}u_{yy})+15u_{yy}(D^{-1}u_{yyyy})+45u_{yy}( D^{-1}u_{yy})^2,\\
&\Big(\frac{D_t^6\{f\cdot f\}}{f^2}\Big)_x=u_{6t}+15u_{tttt}(D^{-1}u_{tt})+15u_{tt}(D^{-1} u_{tttt})+45u_{tt}(D^{-1} u_{tt})^2,\\
&\Big(\frac{D_x^5D_y\{f\cdot f\}}{f^2}\Big)_x=u_{xxxxxy}+10u_xu_{xxxy}+5u_{xxx}u_{xy}+5u_yu_{xxxx}+10u_{xx}u_{xxy}\nonumber\\
&+15u_x^2u_{xy}+30u_xu_yu_{xx},
\end{align}}
{\small\begin{align}
&\Big(\frac{D_x^5D_t\{f\cdot f\}}{f^2}\Big)_x=u_{xxxxxt}+10u_xu_{xxxt}+5u_{xxx}u_{xt}+5u_tu_{xxxx}+10u_{xx}u_{xxt}\nonumber\\
&+15u_x^2u_{xt}+30u_xu_tu_{xx},\\
&\Big(\frac{D_x^4D_y^2\{f\cdot f\}}{f^2}\Big)_x=u_{xxxxyy}+6u_xu_{xxyy}+6u_{xx}u_{xyy}+8u_{xy}u_{xxy}+8u_{y}u_{xxxy}\nonumber\\
&+u_{xxx}u_{yy}+3u_x^2u_{yy}+24u_xu_yu_{xy}+12u_{xx}u_y^2+\Big[6u_xu_{xx}+u_{xxxx}\Big](D^{-1}u_{yy}),\\
&\Big(\frac{D_x^4D_t^2\{f\cdot f\}}{f^2}\Big)_x=u_{xxxxtt}+6u_xu_{xxtt}+6u_{xx}u_{xtt}+8u_{xt}u_{xxt}+8u_{t}u_{xxxt}\nonumber\\
&+u_{xxx}u_{tt}+3u_x^2u_{tt}+24u_xu_tu_{xt}+12u_{xx}u_t^2+\Big[6u_xu_{xx}+u_{xxxx}\Big](D^{-1}u_{tt}),\\
&\Big(\frac{D_x^3D_t^3\{f\cdot f\}}{f^2}\Big)_x=u_{xxxttt}+9u_tu_{xxtt}+3u_{tt}u_{xxt}+3u_{xx}u_{ttt}+9u_{t}u_xu_{tt}\nonumber\\
&+9u_{xt}u_{xtt}+3u_{x}u_{xttt}+18u_t^2u_{xt}+\Big[3u_{xxxt}+9u_tu_{xx}+9u_xu_{xt}\Big](D^{-1}u_{tt}),\\
&\Big(\frac{D_x^3D_y^3\{f\cdot f\}}{f^2}\Big)_x=u_{xxxyyy}+9u_yu_{xxyy}+3u_{yy}u_{xxy}+3u_{xx}u_{yyy}+9u_{y}u_xu_{yy}\nonumber\\
&+9u_{xy}u_{xyy}+3u_{x}u_{xyyy}+18u_y^2u_{xy}+\Big[3u_{xxxy}+9u_yu_{xx}+9u_xu_{xy}\Big](D^{-1}u_{yy}),\\
&\Big(\frac{D_x^2D_t^4\{f\cdot f\}}{f^2}\Big)_x=u_{xxtttt}+8u_tu_{xttt}+6u_{tt}u_{xtt}+8u_{xt}u_{ttt}+12u_t^2u_{tt}\nonumber\\
&+u_{xx}(D^{-1}u_{tttt})+u_xu_{tttt}+\Big[6u_{xxtt}+24u_tu_{xt}+6u_xu_{tt}+3u_{xx}(D^{-1}u_{tt})\Big](D^{-1}u_{tt}),\\
&\Big(\frac{D_x^2D_y^4\{f\cdot f\}}{f^2}\Big)_x=u_{xxyyyy}+8u_yu_{xyyy}+6u_{yy}u_{xyy}+8u_{xy}u_{yyy}+12u_y^2u_{yy}\nonumber\\
&+u_{xx}(D^{-1}u_{yyyy})+u_xu_{yyyy}+\Big[6u_{xxyy}+24u_yu_{xy}+6u_xu_{yy}+3u_{xx}(D^{-1}u_{yy})\Big](D^{-1}u_{yy}),\\
&\Big(\frac{D_y^5D_x\{f\cdot f\}}{f^2}\Big)_x=u_{xyyyyy}+5u_yu_{yyyy}+10u_{yy}u_{yyy}+5u_{xy}(D^{-1}u_{yyyy})\nonumber\\
&+\Big[10u_{xyyy}+30u_yu_{yy}+15u_{xy}(D^{-1}u_{yy})\Big](D^{-1}u_{yy}),\\
&\Big(\frac{D_t^5D_x\{f\cdot f\}}{f^2}\Big)_x=u_{xttttt}+5u_tu_{tttt}+10u_{tt}u_{ttt}+5u_{xt}(D^{-1}u_{tttt})\nonumber\\
&+\Big[10u_{xttt}+30u_tu_{tt}+15u_{xt}(D^{-1}u_{tt})\Big](D^{-1}u_{tt}),\\
&\Big(\frac{D_y^5D_t\{f\cdot f\}}{f^2}\Big)_x=u_{yyyyyt}+10u_{yy}(D^{-1}u_{yyyt})+10u_{yyyt}(D^{-1}u_{yy})+5u_{yt}(D^{-1}u_{yyyy})\nonumber\\
&+5u_{yyyy}(D^{-1}u_{yt})+15u_{yt}(D^{-1}u_{yy})^2+30u_{yy}(D^{-1}u_{yy})(D^{-1}u_{yt}),\\
&\Big(\frac{D_t^5D_y\{f\cdot f\}}{f^2}\Big)_x=u_{ttttty}+10u_{tt}(D^{-1}u_{ttty})+10u_{ttty}(D^{-1}u_{tt})+5u_{yt}(D^{-1}u_{tttt})\nonumber\\
&+5u_{tttt}(D^{-1} u_{yt})+15u_{yt}(D^{-1}u_{tt})^2+30u_{tt}(D^{-1}u_{tt})(D^{-1}u_{yt}),\\
&\Big(\frac{D_y^4D_t^2\{f\cdot f\}}{f^2}\Big)_x=u_{yyyytt}+u_{tt}(D^{-1} u_{yyyy})+u_{yyyy}(D^{-1}u_{tt})+6u_{yy}(D^{-1}u_{tt})(D^{-1}u_{yy})
\nonumber\\
&+24u_{yt}(D^{-1}u_{yt})(D^{-1}u_{yy})
+6u_{yy}(D^{-1}u_{yytt})
+8u_{yt}(D^{-1}u_{yyyt})\nonumber\\
&+8u_{yyyt}(D^{-1}u_{yt})
+6u_{yytt}(D^{-1}u_{yy})+3u_{tt}(D^{-1}u_{yy})^2+2u_{yy}(D^{-1}u_{yt})^2,\\
&\Big(\frac{D_t^4D_y^2\{f\cdot f\}}{f^2}\Big)_x=u_{ttttyy}+u_{yy}(D^{-1} u_{tttt})+u_{tttt}(D^{-1}u_{yy})
+6u_{tt}(D^{-1}u_{yy})(D^{-1}u_{tt})
\nonumber\\
&+24u_{yt}(D^{-1}u_{yt})(D^{-1}u_{tt})
+6u_{tt}(D^{-1}u_{yytt})
+8u_{yt}(D^{-1}u_{ttty})\nonumber\\
&+8u_{ttty}(D^{-1}u_{yt})
+6u_{yytt}(D^{-1}u_{tt})
+3u_{yy}(D^{-1}u_{tt})^2+2u_{tt}(D^{-1}u_{yt})^2,\\
&\Big(\frac{D_y^3D_t^3\{f\cdot f\}}{f^2}\Big)_x=u_{yyyttt}+3u_{yttt}(D^{-1}u_{yy})+9u_{yt}(D^{-1}u_{yytt})+9u_{yytt}(D^{-1}u_{yt})\nonumber\\
&+3u_{tt}(D^{-1}u_{yyyt})+3u_{yyyt}(D^{-1}u_{tt})+18u_{yt}(D^{-1}u_{yt})^2+3u_{yy}(D^{-1}u_{yttt})\nonumber\\
&+9u_{yt}(D^{-1}u_{tt})(D^{-1}u_{yy})+9u_{yy}(D^{-1}u_{yt})(D^{-1}u_{tt})
\nonumber\\&+9u_{tt}(D^{-1}u_{yt})(D^{-1}u_{yy}).
\end{align}}

\vspace{0.5cm}
\noindent
{\bf II) Hirota bilinear forms with triple products}

\vspace{0.5cm}
\noindent
For $n=2$: Hirota bilinear forms of fourth order.
{\small\begin{align}
&\Big(\frac{D_{t}\,D_x^2\,D_y \,\{f\cdot f\}}{f^2}\Big)_x=u_{xxyt}+u_xu_{yt}+2u_yu_{xt}+2u_tu_{xy}+u_{xx}(D^{-1}u_{yt}),\\
&\Big(\frac{D_{t}^2\,D_x\,D_y \,\{f\cdot f\}}{f^2}\Big)_x=u_{xytt}+u_yu_{tt}+2u_tu_{yt}+u_{xy}(D^{-1}u_{tt})+2u_{xt}(D^{-1}u_{yt}),\\
&\Big(\frac{D_{t}\,D_x\,D_y^2 \,\{f\cdot f\}}{f^2}\Big)_x=u_{xtyy}+u_tu_{yy}+2u_yu_{yt}+u_{xt}(D^{-1}u_{yy})+2u_{xy}(D^{-1}u_{yt}).
\end{align}}
\noindent
For $n=3$: Hirota bilinear forms of order six.
{\small\begin{align}
&\Big(\frac{D_{t}\,D_x\,D_y^4 \,\{f\cdot f\}}{f^2}\Big)_x=u_{yyyyxt}+u_tu_{yyyy}+4u_{yt}u_{yyy}+6u_{yy}u_{tyy}+4u_yu_{tyyy}\nonumber\\
&+\Big(6u_tu_{yy}+12u_yu_{yt}+6u_{xtyy}+3u_{xt}(D^{-1}u_{yy})\Big)(D^{-1}u_{yy})+4u_{xy}(D^{-1}u_{tyyy})\nonumber\\
&+u_{xt}(D^{-1}u_{yyyy})+\Big(4u_{xyyy}+12u_yu_{yy}+12u_{xy}(D^{-1}u_{yy})\Big)(D^{-1}u_{yt}),\\
&\Big(\frac{D_{t}\,D_x^2\,D_y^3 \,\{f\cdot f\}}{f^2}\Big)_x=u_{yyyxxt}+u_xu_{yyyt}+6u_yu_{xyyt}+6u_y^2u_{yt}+2u_tu_{xyyy}+6u_tu_yu_{yy}\nonumber\\
&+6u_{xy}u_{yyt}+3u_{yt}u_{xyy}+2u_{yyy}u_{xt}+\Big(3u_xu_{yt}+6u_yu_{xt}+6u_tu_{xy}+3u_{xxyt}\Big)(D^{-1}u_{yy})\nonumber\\
&+3u_{yy}u_{xyt}+\Big(3u_{xxyy}+12u_yu_{xy}+3u_xu_{yy}+3u_{xx}(D^{-1}u_{yy})\Big)(D^{-1}u_{yt})+u_{xx}(D^{-1}u_{yyyt}),\\
&\Big(\frac{D_{t}\,D_x^3\,D_y^2 \,\{f\cdot f\}}{f^2}\Big)_x=u_{xxxyyt}+6u_{xy}u_{xyt}+3u_{xx}u_{yyt}+2u_{yt}u_{xxy}+3u_{xt}u_{xyy}+6u_yu_{xxyt}\nonumber\\
&+3u_xu_{xyyt}+6u_{xt}u_y^2+3u_tu_{xxyy}+3u_xu_tu_{yy}+u_{yy}u_{xxt}+12u_tu_yu_{xy}+6u_xu_yu_{yt}\nonumber\\
&+\Big(2u_{xxxy}+6u_xu_{xy}+6u_yu_{xx}\Big)(D^{-1}u_{yt})+\Big(u_{xxxt}+3u_xu_{xt}+3u_tu_{xx}\Big)(D^{-1}u_{yy}),\\
&\Big(\frac{D_{t}\,D_x^4\,D_y \,\{f\cdot f\}}{f^2}\Big)_x=u_{xxxxyt}+12u_tu_yu_{xx}+u_{xxx}u_{yt}+12u_xu_yu_{xt}+12u_xu_tu_{xy}\nonumber\\
&+4u_{xy}u_{xxt}+6u_{xx}u_{xyt}+4u_{xt}u_{xxy}+4u_yu_{xxxt}+6u_xu_{xxyt}+3u_x^2u_{yt}+4u_tu_{xxxy}\nonumber\\
&+\Big(u_{xxxx}+6u_xu_{xx}\Big)(D^{-1}u_{yt}),\\
&\Big(\frac{D_{x}\,D_t^2\,D_y^3 \,\{f\cdot f\}}{f^2}\Big)_x=u_{yyyxtt}+3u_{ytt}u_{yy}+6u_{yt}u_{yyt}+3u_yu_{yytt}+2u_tu_{yyyt}+u_{tt}u_{yyy}\nonumber\\
&+\Big(3u_{xytt}+3u_{y}u_{tt}+6u_tu_{yt}+6u_{xt}(D^{-1}u_{yt})\Big)(D^{-1}u_{yy})+3u_{xy}(D^{-1}u_{yytt})\nonumber\\
&+\Big(u_{xyyy}+3u_yu_{yy}+3u_{xy}(D^{-1}u_{yy})\Big)(D^{-1}u_{tt})+2u_{xt}(D^{-1}u_{yyyt})\nonumber\\
&+\Big(6u_{xyyt}+6u_tu_{yy}+12u_yu_{yt}+6u_{xy}(D^{-1}u_{yt})\Big)(D^{-1}u_{yt}),\\
&\Big(\frac{D_{x}\,D_t^4\,D_y \,\{f\cdot f\}}{f^2}\Big)_x=u_{ttttxy}+u_yu_{tttt}+4u_{yt}u_{ttt}+6u_{tt}u_{ytt}+4u_tu_{yttt}+u_{xy}(D^{-1}u_{tttt})\nonumber\\
&+\Big(6u_yu_{tt}+12u_tu_{yt}+6u_{xytt}+3u_{xy}(D^{-1}u_{tt})\Big)(D^{-1}u_{tt})+4u_{xt}(D^{-1}u_{yttt})\nonumber\\
&+\Big(4u_{xttt}+12u_tu_{tt}+12u_{xt}(D^{-1}u_{tt})\Big)(D^{-1}u_{yt}),\\
&\Big(\frac{D_{x}\,D_t^3\,D_y^2 \,\{f\cdot f\}}{f^2}\Big)_x=u_{tttxyy}+3u_{tyy}u_{tt}+6u_{yt}u_{tty}+3u_tu_{yytt}+2u_yu_{ttty}+u_{yy}u_{ttt}\nonumber\\
&+\Big(3u_{xtyy}+3u_{t}u_{yy}+6u_yu_{yt}+6u_{xy}(D^{-1}u_{yt})\Big)(D^{-1}u_{tt})+3u_{xt}(D^{-1}u_{yytt})\nonumber\\
&+\Big(u_{xttt}+3u_tu_{tt}+3u_{xt}(D^{-1}u_{tt})\Big)(D^{-1}u_{yy})+2u_{xy}(D^{-1}u_{ttty})\nonumber\\
&+\Big(6u_{xtty}+6u_yu_{tt}+12u_tu_{yt}+6u_{xt}(D^{-1}u_{yt})\Big)(D^{-1}u_{yt}),\\
&\Big(\frac{D_{y}\,D_x^2\,D_t^3 \,\{f\cdot f\}}{f^2}\Big)_x=u_{tttxxy}+u_xu_{ttty}+6u_tu_{xtty}+6u_t^2u_{yt}+2u_yu_{xttt}+6u_tu_yu_{tt}+3u_{tt}u_{xyt}\nonumber\\
&+6u_{xt}u_{tty}+3u_{yt}u_{xtt}+2u_{ttt}u_{xy}+\Big(3u_xu_{yt}+6u_tu_{xy}+3u_{xxyt}+6u_yu_{xt}\Big)(D^{-1}u_{tt})\nonumber\\
&+\Big(3u_{xxtt}+3u_xu_{tt}+12u_tu_{xt}+3u_{xx}(D^{-1}u_{tt})\Big)(D^{-1}u_{yt})+u_{xx}(D^{-1}u_{ttty}),\\
&\Big(\frac{D_{y}\,D_x^3\,D_t^2 \,\{f\cdot f\}}{f^2}\Big)_x=u_{xxxtty}+6u_{xt}u_{xyt}+3u_{xx}u_{tty}+2u_{yt}u_{xxt}+3u_{xy}u_{xtt}+6u_tu_{xxyt}\nonumber\\
&+3u_xu_{xtty}+6u_{xy}u_t^2+3u_yu_{xxtt}+3u_xu_yu_{tt}+u_{tt}u_{xxy}+12u_tu_yu_{xt}+6u_xu_tu_{yt}\nonumber\\
&+\Big(2u_{xxxt}+6u_xu_{xt}+6u_tu_{xx}\Big)(D^{-1}u_{yt})+\Big(u_{xxxy}+3u_xu_{xy}+3u_yu_{xx}\Big)(D^{-1}u_{tt}),
\end{align}}
{\small\begin{align}
&\Big(\frac{D_{x}^2\,D_t^2\,D_y^2 \,\{f\cdot f\}}{f^2}\Big)_x=u_{xxyytt}+8u_tu_yu_{yt}+u_{yy}u_{xtt}+u_xu_{yytt}+u_{tt}u_{yyx}+4u_{xt}u_{yyt}
\nonumber\\
&+4u_yu_{xytt}+4u_tu_{yyxt}+4u_{yx}u_{ytt}+2u_y^2u_{tt}+2u_t^2u_{yy}+4u_{yt}u_{xyt}\nonumber\\
&+u_{xx}(D^{-1}u_{yytt})+\Big(4u_yu_{xy}+u_xu_{yy}+u_{yyxx}\Big)(D^{-1}u_{tt})\nonumber\\
&+\Big(4u_tu_{xt}+u_xu_{tt}+u_{xxtt}+u_{xx}(D^{-1}u_{tt})\Big)(D^{-1}u_{yy})\nonumber\\
&+\Big(8u_yu_{xt}+4u_xu_{yt}+8u_tu_{xy}+4u_{yxxt}+2u_{xx}(D^{-1} u_{yt})\Big)(D^{-1}u_{yt}).
\end{align}}

\section{Acknowledgment}
  This work is partially supported by the Scientific
and Technological Research Council of Turkey (T\"{U}B\.{I}TAK).\\

\end{document}